\documentclass[nofootinbib, aps,12pt,preprintnumbers]{revtex4-1}

\usepackage{graphicx}
\usepackage{amssymb}
\usepackage{textcomp}
\usepackage{amsmath}
\usepackage{bm}
\usepackage{times}
\usepackage{epsfig}
\usepackage{color}
\usepackage{array,multirow}

\newcommand{\Tr}{\mbox{Tr$\;$}}
\newcommand{\susy}{{\mbox{\scriptsize SUSY}}}

\newcommand{\trix}[1]{\left(\begin{array}{#1}}
\newcommand{\notrix}{\end{array}\right)}
\newcommand{\comment}[1]{}

\def\beq{\begin{equation}}
\def\eeq{\end{equation}}
\def\bea{\begin{eqnarray}}
\def\eea{\end{eqnarray}}

\begin{document}
\title{\Large  {{\bf{The Minimal SUSY $B-L$ Model: Simultaneous Wilson Lines \\
and String Thresholds}}}}
\author{{Rehan Deen$^{1}$, Burt A.~Ovrut$^{1}$ and Austin Purves$^{1,2}$} \\[2mm]
    {\it $^{1}$ Department of Physics, University of Pennsylvania} \\
   {\it Philadelphia, PA 19104--6396}\\
   {\it $^{2}$ Department of Physics, Manhattanville College}\\
   {\it Purchase, NY 10577} \\[4mm]
}
\date{\today}

{\let\thefootnote\relax\footnotetext{\mbox{ovrut@elcapitan.hep.upenn.edu,  ~austin.purves@mville.edu, ~rdeen@sas.upenn.edu~} }}
\setcounter{footnote}{0}

\begin{abstract}
\ \\ [10mm]
{\bf ABSTRACT:} In previous work, we presented a statistical scan over the soft supersymmetry breaking parameters of the minimal SUSY $B-L$ model. For specificity of calculation, unification of the gauge parameters was enforced by allowing the two  ${\mathbb Z}_{3}\times {\mathbb Z}_{3}$ Wilson lines to have mass scales separated by approximately an order of magnitude. This introduced an additional ``left-right'' sector below the unification scale. In this paper, for three important reasons, we modify our previous analysis by demanding that the mass scales of the two Wilson lines be simultaneous and equal to an ``average unification'' mass $\left<M_{U}\right>$. The present analysis is 1) more ``natural'' than the previous calculations, which were only valid in a very specific region of the Calabi-Yau moduli space, 2) the theory is conceptually simpler in that the left-right sector has been removed and 3) in the present analysis the lack of gauge unification is due to threshold effects--particularly heavy string thresholds, which we calculate statistically in detail.
As in our previous work, the theory is renormalization group evolved from $\left<M_{U}\right>$ to the electroweak scale--being subjected, sequentially, to the requirement of radiative $B-L$ and electroweak symmetry breaking, the present experimental lower bounds on the $B-L$ vector boson and sparticle masses, as well as the lightest neutral Higgs mass of $\sim$125 GeV. The subspace of soft supersymmetry breaking masses that satisfies all such constraints is presented and shown to be substantial.

\end{abstract}

\maketitle

\tableofcontents

%
\section{Introduction}
%

Within the context of the heterotic superstring and heterotic M-theory \cite{Lukas:1997fg, Lukas:1998yy, Lukas:1998tt, Donagi:1999ez}, there have been a number of vacuum states whose four-dimensional low energy effective field theory \cite{Ovrut:1979pk} has the exact spectrum of the MSSM--with or without right-handed neutrino chiral multiplets--and, to prohibit rapid proton decay, contains R-parity \cite{Font:1989ai,Martin:1997ns,Martin:1992mq}--either as a discrete symmetry or as a subgroup of an anomaly free  $U(1)$ extension of the standard model gauge group \cite{Krauss:1988zc,Aulakh:1999cd, Aulakh:2000sn, Babu:2008ep, Feldman:2011ms, FileviezPerez:2011dg,Aulakh:1982yn, Hayashi:1984rd, Mohapatra:1986aw, Masiero:1990uj}. One such vacuum was presented in \cite{Braun:2005zv, Braun:2005nv, Ambroso:2009jd, Ambroso:2010pe} and will be referred to as the $B-L$ MSSM. Be that as it may, this is only the first step in finding a realistic heterotic string vacuum. Any such theory must also also be compatible with all presently observed low-energy phenomenology; that is, it must spontaneously break electroweak (EW) symmetry at the observed scale, must be compatible with the newly discovered Higgs particle with mass $\sim 125$~GeV \cite{Aad:2014aba, Chatrchyan:2013lba}, have all sparticle masses above the present observational lower bounds and--assuming R-parity is contained in an additional $U(1)$ symmetry--spontaneously break that Abelian group with an associated gauge boson mass in excess of the present experimental lower bound.

In a series of papers \cite{Ovrut:2012wg, Marshall:2014kea, Marshall:2014cwa, Ovrut:2014rba, Ovrut:2015uea}, the $B-L$ MSSM was examined in detail--using a random statistical sampling of the initial set of soft  supersymmetry (SUSY) breaking parameters--and the results confronted with these  phenomenological requirements. It was shown, {\it within a restricted region of the compactification moduli space,} that the $B-L$ MSSM easily passed each of these requirements for a large and basically uncorrelated set of initial conditions.
Furthermore, this analysis led to a series of low energy predictions--for example, directly relating the lightest stop decay channels and branching ratios to the neutrino mass hierarchy and mixing angles--thus linking LHC experimental results to neutrino measurements. That is, the $B-L$ MSSM is a possible candidate for a phenomenologically acceptable theory of the real world--a statement that will be directly testable as its structure and predictions are confronted with upcoming data from the LHC, neutrino experiments and cosmological observations. For these reasons, in this paper we extend the results of \cite{Ovrut:2012wg, Marshall:2014kea, Marshall:2014cwa, Ovrut:2014rba, Ovrut:2015uea} to a {\it more general, and more natural, region} of the $B-L$ MSSM moduli space.

To quantify this, we should briefly discuss the structure of the $B-L$ MSSM. It arises in the observable sector of heterotic M-theory compactified to four-dimensions on a Shoen Calabi-Yau (CY) threefold \cite{Braun:2004xv} with first homotopy group $\pi_{1}={\mathbb{Z}}_{3} \times {\mathbb{Z}}_{3}$. This manifold admits a specific slope-stable holomorphic vector bundle \cite{Donagi:2000zf} with structure group $SU(4)\subset E_{8}$, as well as two Wilson lines--each wrapped over a two-cycle associated with a different ${\mathbb{Z}}_{3}$ homotopy factor. This theory has three, in principle distinct, mass scales--$M_{U}$ at which the gauge bundle spontaneously breaks $E_{8}$ to $SO(10)$ and two Wilson line mass scales, which we denote by $M_{\chi_{3R}}$ and $M_{\chi_{B-L}}$, associated with the inverse radii of their respective two-cycles. In our previous analysis, we worked {\it in a restricted region of CY moduli space} where the radius of one two-cycle is distinctly smaller than that of the other; that is, we chose $M_{U} \simeq M_{\chi_{B-L}} > M_{\chi_{3R}}$. By taking the scale of separation of the two Wilson lines to be approximately an order of magnitude, one can exactly unify all gauge coupling parameters-- thus specifying boundary conditions in the renormalization group equations (RGEs). However, this specificity comes at the cost of introducing an {\it additional scaling regime}. Between $M_{\chi_{B-L}}$ and $M_{\chi_{3R}}$ the effective theory is that of the ``left-right'' model \cite{Ovrut:2012wg,Babu:2008ep,FileviezPerez:2008sx,Hayashi:1984rd} with gauge group $SU(3)_{C} \times SU(2)_{L} \times SU(2)_{R} \times U(1)_{B-L}$ and a specific particle spectrum that can be computed from string theory. It is only for energy-momentum below the lightest Wilson line mass $M_{\chi_{3R}}$ that one obtains the spectrum and $SU(3)_{C} \times SU(2)_{L} \times U(1)_{Y} \times U(1)_{B-L}$ gauge symmetry of the $B-L$ MSSM.  

In this paper, we generalize and simplify the phenomenological analysis of the $B-L$ MSSM by working in a {\it generic} region of CY moduli space where the radii of the two Wilson lines and the average radius of the CY manifold are all approximately equal: that is,  with $M_{U} \simeq M_{\chi_{B-L}} \simeq M_{\chi_{3R}}$. This generalization is significant in that 1) {\it the region of moduli space is much larger and more ``natural''} than that used previously and 2) {\it the ``left-right'' scaling region is eliminated}, with the $B-L$ MSSM emerging immediately below the compactification scale--thus simplifying the scaling regimes. Of course, the four $B-L$ MSSM gauge couplings will no longer unify near the scale of the CY radius. This does somewhat complicate the RG analysis. However, it opens the door for a discussion of unification of all gauge couplings with the gravitational coupling at the ``string scale''--as has been discussed by many authors in \cite{Kaplunovsky:1992vs, Kaplunovsky:1995jw, Mayr:1993kn, Dienes:1995sq, Dienes:1996du, Kiritsis:1996dn, Dolan:1992nf, Nilles:1997vk, Nilles:1998uy, Ghilencea:2001qq, Klaput:2010dg, deAlwis:2012bm, Bailin:2014nna}. More specifically, such unification should take place at tree level. However, at the one-loop (and higher) level one expects such unification to be split by ``threshold'' corrections. These are due to several effects, such as the inclusion of field theory thresholds at each of the SUSY, $B-L$ and ``unification'' scales, 
and genus-one string theory corrections. Since in this analysis the latter is expected to be the largest, we will focus exclusively on them. By running the four $B-L$ MSSM gauge parameters up to the string scale, we will 3) {\it statistically compute the heavy string threshold corrections for each gauge coupling}. Furthermore, we will statistically compute the hypercharge gauge threshold and, by subtracting various thresholds, analyze the moduli dependent sub-component of each. Finally, there is yet another important benefit of analyzing the $B-L$ MSSM at the generic region of its CY moduli space--although we will not pursue this in the present paper. In our previous work and the present analysis, the scale of the soft SUSY breaking parameters is chosen to be in the TeV region. This is done in order for our low energy phenomenological predictions to be LHC accessible. However, unlike in the case of unified gauge couplings enforced by splitting the Wilson line masses discussed in \cite{Ovrut:2012wg, Marshall:2014kea, Marshall:2014cwa, Ovrut:2014rba, Ovrut:2015uea}--which, for reasons elucidated in the Conclusion, is essentially restricted to the TeV region--for the simultaneous Wilson line masses discussed in this paper, the SUSY breaking mass scale can be taken to be arbitrarily large. This has a number of important applications, both in particle phenomenology and in early universe cosmology \cite{Lukas:1998qs}. We will pursue this in future work \cite{inPreparation}.

The present paper is structured as follows. In Section II we review the salient parts of the $B-L$ MSSM theory, presenting the spectrum, the supersymmetric and the soft SUSY breaking Lagrangians, discussing the generic structure of spontaneous $B-L$ and EW symmetry breaking and setting our notation. Section III is devoted to defining the exact meaning of the ``simultaneous'' Wilson line analysis presented in this paper--as opposed to the ``split'' Wilson line approach in our previous work. It then discusses, in detail, the four relevant mass scales  from ``unification'' to electroweak symmetry breaking. The statistical definitions of the ``unification'' mass and gauge coupling are given in our present context. In Section IV, the three scaling regimes--for both the ``right-side-up'' and ``upside-down'' scenarios--along with the associated gauge coupling beta function parameters are presented. The RG running of the Yukawa couplings, including their transition at the SUSY scale, is discussed. Section V gives a brief review of the ``statistical'' approach to setting the initial soft supersymmetry breaking parameters at the ``unification'' scale presented in detail in our previous work. In Section VI, the experimental constraints on the sparticle masses, the heavy vector boson mass and the lightest neutral Higgs mass are presented. We then solve the RGEs--for randomly chosen initial soft SUSY breaking parameters-- sequentially from the ``unification'' scale down through the EW breaking scale subject to these constraints. Plots--and the exact number--of the initial points that sequentially satisfy the experimental constraints are given; ending with the robust number of phenomenologically ``valid'' black points that satisfy all experimental constraints. A brief analysis of both the LSP and non-LSP spectra is then given. In Section VII, we briefly discuss fine-tuning in the $B-L$ MSSM. In Section VIII, we statistically calculate the heavy string threshold corrections for each of the four $B-L$ MSSM gauge couplings, as well as the hypercharge gauge threshold, and analyze the moduli dependent differences of these quantities. Finally, in Section IX, we present our conclusions.

%
\section{The Minimal SUSY $B-L$ Model}
\label{sec:model}
%
In this section, we briefly review the minimal anomaly free extension of the MSSM with gauge group
\begin{eqnarray}
	SU(3)_C\times SU(2)_L\times U(1)_{3R}\times U(1)_{B-L} \ ,
	\label{eq:458}
\end{eqnarray}
whose structure was motivated by heterotic string theory in \cite{Braun:2005nv} and by phenomenological considerations in \cite{Barger:2008wn,Everett:2009vy}. 
Although this model has been discussed in our previous papers \cite{Ovrut:2012wg, Marshall:2014kea, Marshall:2014cwa, Ovrut:2014rba, Ovrut:2015uea}, we outline its main features in this section for specificity and to set our notation.
The Abelian gauge factors $U(1)_{3R}\times U(1)_{B-L}$ can be rotated into physically equivalent charge bases, such as $U(1)_Y\times U(1)_{B-L}$. However, as shown in~\cite{Ovrut:2012wg}, this comes at the cost of introducing kinetic mixing between the gauge fields. We therefore prefer to work in the basis $U(1)_{3R}\times U(1)_{B-L}$. The gauge covariant derivative is
\beq
D=\partial-ig_{3R}I_{3R}W_{3R}-ig_{BL}\frac{I_{BL}}{2}B^\prime \ ,
\eeq
where $I_{3R}$, $I_{BL}$ and $g_{3R}$, $g_{BL}$ are the generators and couplings for the $U(1)_{3R}$ and $U(1)_{B-L}$ groups respectively. The gauge boson associated with $U(1)_{B-L}$ is denoted $B^\prime$ to distinguish it from the gauge boson associated with $U(1)_Y$, which is normally denoted $B$. The factor of $\frac{1}{2}$ in the last term is introduced by redefining the gauge coupling $g_{BL}$, thus simplifying many equations. A radiatively induced vacuum expectation value (VEV) of a right-handed sneutrino will break the Abelian factors $U(1)_{3R} \times U(1)_{B-L}$ to $U(1)_Y$, in analogy with the way the MSSM Higgs fields break $SU(2)_L\times U(1)_Y$ to $U(1)_{EM}$. This process is referred to as ``$B-L$'' symmetry breaking, although technically it breaks a specific combination of the groups generated from $I_{3R}$ and $I_{BL}$, leaving invariant the usual hypercharge group generated by
\beq
	Y = I_{3R} + \frac{I_{BL}}{2} \ . 
\eeq

The particle content of the model is simply that of the MSSM plus three right-handed neutrino chiral multiplets. This amounts to three generations of matter superfields
\begin{eqnarray}
	Q=\trix{c}u\\d\notrix\sim({\bf 3}, {\bf 2}, 0, \frac{1}{3}) & \begin{array}{rl}u^c\sim&(\bar{\bf 3}, {\bf 1}, -1/2, -\frac{1}{3}) \\
	d^c\sim&(\bar{\bf 3}, {\bf 1}, 1/2, -\frac{1}{3})\end{array} \ , \nonumber\\
	L=\trix{c}\nu\\e\notrix\sim({\bf 1}, {\bf 2}, 0, -1)&\begin{array}{rl}\nu^c\sim&({\bf 1}, {\bf 1}, -1/2, 1)\\
	e^c\sim&({\bf 1}, {\bf 1}, 1/2, 1)\end{array} \ ,
	\label{eq:246}
\end{eqnarray}
along with the usual two Higgs supermultiplets
\begin{eqnarray}
	H_u=\trix{c}H_u^+\\H_u^0\notrix&\sim&({\bf 1}, {\bf 2}, 1/2, 0) \ ,\nonumber\\
	H_d=\trix{c}H_d^0\\H_d^-\notrix&\sim&({\bf 1}, {\bf 2}, -1/2, 0) 
	\label{eq:247}
\end{eqnarray}
where we have displayed their $SU(3)_{C}\times SU(2)_{L} \times U(1)_{3R} \times U(1)_{B-L}$ quantum numbers. The superpotential of the $B-L$ MSSM is given by
\begin{eqnarray}
	W=Y_u Q H_u u^c - Y_d Q H_d d^c -Y_e L H_d e^c +Y_\nu L H_u \nu^c+\mu H_u H_d \ ,
\end{eqnarray}
where both generational and gauge indices have been suppressed. In principle, the Yukawa couplings are three-by-three complex matrices. However, the observed smallness of the CKM mixing angles and CP-violating phase imply that the quark Yukawa matrices can be approximated as diagonal and real for the purposes of RG evolution in this paper. The charged lepton Yukawa coupling can be made diagonal and real by moving the PMNS angles and phases into the neutrino Yukawa couplings. The small size of neutrino masses implies that the neutrino Yukawa couplings can be neglected for the purposes of RG evolution in this paper. The smallness of first- and second-generation fermion masses implies that first and second-generation Yukawa quark and charged lepton Yukawa couplings can also be neglected. The $\mu$-parameter can be chosen to be real without loss of generality.

The soft supersymmetry breaking Lagrangian is
\begin{align}
\begin{split}
	-\mathcal L_{\mbox{\scriptsize soft}}  = &
	\left(
		\frac{1}{2} M_3 \tilde g^2+ \frac{1}{2} M_2 \tilde W^2+ \frac{1}{2} M_R \tilde W_R^2+\frac{1}{2} M_{BL} \tilde {B^\prime}^2
	\right.
		\\
	& \left.
		\hspace{0.4cm} +a_u \tilde Q H_u \tilde u^c - a_d \tilde Q H_d \tilde d^c - a_e \tilde L H_d \tilde e^c
		+ a_\nu \tilde L H_u \tilde \nu^c + b H_u H_d + h.c.
	\right)
	\\
	& + m_{\tilde Q}^2|\tilde Q|^2+m_{\tilde u^c}^2|\tilde u^c|^2+m_{\tilde d^c}^2|\tilde d^c|^2+m_{\tilde L}^2|\tilde L|^2
	+m_{\tilde \nu^c}^2|\tilde \nu^c|^2+m_{\tilde e^c}^2|\tilde e^c|^2 \\
	&+m_{H_u}^2|H_u|^2+m_{H_d}^2|H_d|^2 \ ,
\end{split}
\label{home5}
\end{align}
where generation and gauge indices have been suppressed. The $a$-parameters and sfermion soft-mass terms can, in principle, be hermitian matrices in family space. However, this tends to lead to unobserved CP violation. Therefore, we proceed assuming that they are diagonal and real. Furthermore, as discussed in Section V, we assume that the a-parameters are proportional to the Yukawa couplings. This implies that all the $a$-parameters can be neglected except for the (3,3) component of the quark and charged lepton $a$-parameters. The $b$-parameter can be chosen to be both real and positive without loss of generality. Although the gaugino soft masses can be complex in principle, this tends to lead to unobserved flavor and CP violation. Therefore, we proceed assuming that they are real.

The $B-L$ symmetry is spontaneously broken by the VEV in a right-handed sneutrino, which carries the appropriate $I_{3R}$ and $I_{B-L}$ charges to break those symmetries while preserving hypercharge symmetry. This VEV is brought about by a sneutrino soft-mass term becoming tachyonic\footnote{Throughout this paper, we use the term ``tachyon'' to describe a scalar particle with a negative mass squared parameter.} at the TeV scale due to the RGE evolution. As discussed in~\cite{Mohapatra:1986aw, Ghosh:2010hy, Barger:2010iv}, this VEV will be purely in one of the three right-handed sneutrino generations -- not in a linear combination of them. Furthermore, the three generations of right-handed sneutrinoes can be relabeled without loss of generality. Therefore, we henceforth assume that it is the third-generation right-handed sneutrino that acquires a VEV. Electroweak symmetry is broken by VEVs in the neutral components of the up and down Higgs multiplets. The electroweak breaking VEVs and the $B-L$ breaking VEV together lead to small VEVs in all three generations of left-handed sneutrinos.  The above VEVs will be denoted by
\beq
	\left< \tilde \nu^c_3 \right> \equiv \frac{1}{\sqrt 2} v_R, \ \ \left<\tilde \nu_i\right> \equiv \frac{1}{\sqrt 2} {v_L}_i, \ \
	\left< H_u^0\right> \equiv \frac{1}{\sqrt 2}v_u, \ \ \left< H_d^0\right> \equiv \frac{1}{\sqrt 2}v_d,
\eeq
where $i=1,2,3$ is the generation index.

The neutral gauge boson that becomes massive due to $B-L$ symmetry breaking is referred to as $Z_R$. Defining $v^{2}=v_{u}^{2}+v_{d}^{2}$, and assuming that $v^{2} \ll v_{R}^{2}$,  $Z_{R}$ acquires to leading order a mass of
\beq
	M_{Z_R}^2 = \frac{1}{4}\left(g_{3R}^2+g_{BL}^2 \right) v_R^2\ .
	\label{eq:237}
\eeq
The hypercharge gauge coupling is given by
\beq
	\label{eq:Y.3R.BL}
	g_Y = g_{3R} \sin \theta_R = g_{BL}\cos \theta_R \ ,
\eeq
where
\beq
		\cos \theta_R = \frac{g_{3R}}{\sqrt{g_{3R}^2+g_{BL}^2}} \ .
\eeq

The smallness of the neutrino masses implies, first, that the neutrino Yukawa couplings are small and, second, that the  left-handed sneutrino VEVs are much smaller than the electroweak scale. In this limit, the minimization conditions of the potential simplify to 
\begin{align}
	\label{eq:MC.vR}
         v_R^2=&\frac{-8m^2_{\tilde \nu_{3}^c}  + g_{3R}^2\left(v_u^2 - v_d^2 \right)}{g_{3R}^2+g_{BL}^2} \ , 
	\\
	{v_L}_i=&\frac{\frac{v_R}{\sqrt 2}(Y_{\nu_{i3}}^* \mu v_d-a_{\nu_{i3}}^* v_u)}
			{m_{\tilde L_{i}}^2-\frac{g_2^2}{8}(v_u^2-v_d^2)-\frac{g_{BL}^2}{8}v_R^2} \ ,
	\\
	\label{eq:EW.mu}
	\frac{1}{2} M_Z^2 =&-\mu^2+\frac{m_{H_u}^2\tan^2\beta-m_{H_d}^2}{1-\tan^2\beta} \ ,
	\\
	\label{eq:EW.b}
	\frac{2b}{\sin2\beta}=&2\mu^2+m_{H_u}^2+m_{H_d}^2 \ .
\end{align}
Noting from above that $ |v_{u}^{2}-v_{d}^{2}|\ll|m_{\tilde\nu_3^c}^{2}|$, equations \eqref{eq:237} and \eqref{eq:MC.vR} can be combined to give
\beq
	\label{eq:MZR.mnuc}
	M_{Z_R}^2 = -2 m_{\tilde \nu^c_3}^2\ .
\eeq

The VEV in the third-generation right-handed sneutrino induces spontaneous bilinear R-parity violation through the operators
\begin{equation}
	\label{W.brpv}
	W \supset \epsilon_i \,  L_i \,  H_u - \frac{1}{\sqrt 2 }{Y_e}_i \, {v_L}_i \,  H_d^- \,   e^c_i \ ,
\end{equation}
where
\begin{equation}
	\epsilon_i  \equiv \frac{1}{\sqrt 2} {Y_\nu}_{i3} v_R \ .
\end{equation}
Bilinear R-parity violation has been discussed extensively, including its relevance to neutrino masses. See, for example, some early works~\cite{Mukhopadhyaya:1998xj,Chun:1998ub, Chun:1999bq, Hirsch:2000ef}. The Lagrangian of this model contains additional bilinear terms due to the sneutrino VEVs:
\begin{align}
\begin{split}
	\label{L.n}
	\mathcal{L} \supset &
	- \frac{1}{2}{v_L}_i^* \left[ g_2 \left(\sqrt 2 \, e_i \tilde W^+ 
	+  \nu_i \tilde W^0\right) - g_{BL} \nu_i \tilde B' \right]
	\\
	&
	-\frac{1}{2} v_R \left[-g_R \nu_3^c \tilde W_R + g_{BL} \nu_3^c \tilde B' \right]+ \text{h.c.}
\end{split}
\end{align}
The R-parity violating terms in this model have a variety of interesting consequences that have been studied in a number of different contexts. These include LHC studies~\cite{Barger:2008wn, Everett:2009vy, FileviezPerez:2012mj, Perez:2013kla}, predictions for neutrinos~\cite{Mohapatra:1986aw, Ghosh:2010hy, Barger:2010iv}, and connections between the two~\cite{Marshall:2014kea, Marshall:2014cwa}. It has been shown that the R-parity violation can give rise to Majorana neutrino masses, with the lightest left-handed neutrino being massless. There is also a pair of sterile right-handed neutrinos that can have cosmological implications~\cite{Perez:2013kla}.

The minimal supersymmetric $B-L$ model, reviewed in this section, will be referred to simply as the $B-L$ MSSM throughout the rest of this paper. We now turn to connecting the phenomenology of the $B-L$ MSSM to its high-scale origins. Specifically, we are considering the possibility that the $B-L$ MSSM is the observable sector  of the low-energy effective theory of an $E_8\times E_8$ heterotic string theory. In this context, the $B-L$ MSSM gauge group unifies into an $SO(10)$ gauge group, which is itself the commutant of the $SU(4)$ structure group of the observable sector $E_{8}$ vector bundle on the CY threefold. We have previously studied the $B-L$ MSSM in this context~\cite{Ovrut:2015uea}. In this paper, however, we will study the effects of string threshold corrections on gauge unification. This requires a discussion of gauge unification--to which we now turn.

%
\section{Journey From the ``Unification'' Scale}
\label{sec:unif}
%
This section outlines the scales and scaling regimes associated with the evolution of the $B-L$ MSSM from ``unification'' to the electroweak scale. Compactification to four dimensions yields a unified gauge group, $SO(10)$, at mass scale $M_{U}$. This unified gauge group is broken by two Abelian Wilson lines, denoted by $\chi_{3R}$ and $\chi_{B-L}$. The mass scales associated with these Wilson lines, $M_{\chi_{3R}}$ and $M_{\chi_{B-L}}$ respectively, depend on the inverse radii of the 2-cycles over which they are wrapped. These, in turn, depend on the chosen point in the CY moduli space. Generically, one expects that the two Wilson line masses are approximately the same and close to the $SO(10)$ unification scale. That is, one ``naturally'' expects
\begin{equation}
 M_{U} \simeq  M_{\chi_{B-L}}  \simeq  M_{\chi_{3R}}
 \label{s1}
 \end{equation}
over a wide region of the CY moduli space. However, as one moves away from these generic points the Wilson line mass scales need not remain the same. This leads to an intermediate regime between the two scales associated with the Wilson lines. The particle content and gauge group in this intermediate regime depends on which Wilson line has a higher associated mass. If $M_{U} \simeq M_{\chi_{B-L}} > M_{\chi_{3R}}$, the particle content and gauge group of the intermediate regime is that of a ``left-right'' model. If $M_{U} \simeq M_{\chi_{3R}} > M_{\chi_{B-L}}$, the particle content and gauge group of the intermediate regime is similar to that of a``Pati-Salam'' model.

\begin{figure}
\centering
	\includegraphics[scale=1.2]{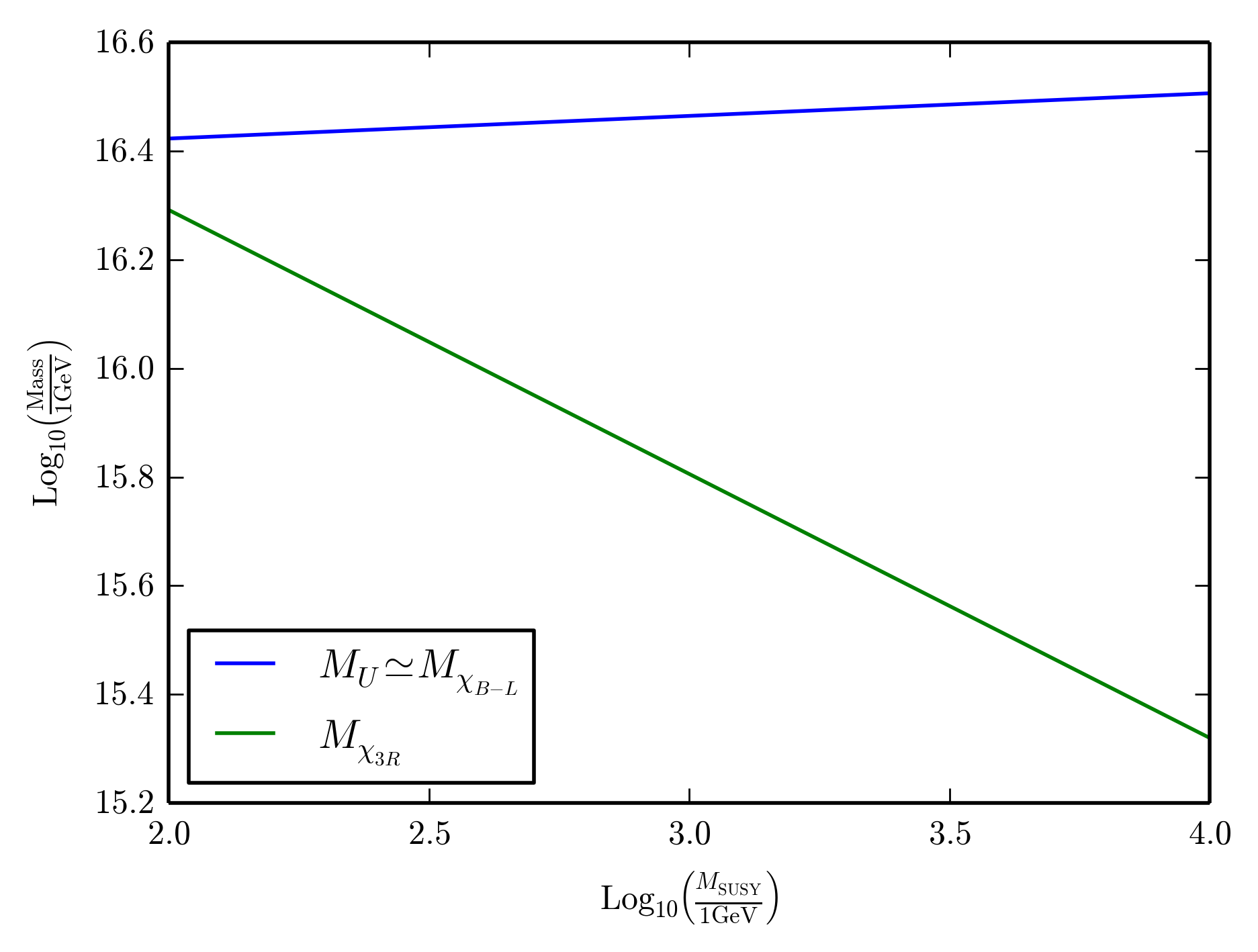}
	\caption{The $M_{U} \simeq M_{\chi_{B-L}}$ unification mass and $M_{\chi_{3R}}$ as functions of the SUSY scale in the ``left-right'' scenario.}
	\label{fig:MU.MI.MSUSY}
\end{figure}

In each case, the lower-mass Wilson line breaks the model in the intermediate regime to the $B-L$ MSSM.
In fact, it was shown in~\cite{Ovrut:2012wg} that exact gauge coupling unification at one-loop {\it requires} that these scales be different. For specificity of the RGE calculation, it was convenient to {\it impose} precise gauge coupling unification. Hence, in~\cite{Ovrut:2012wg} we studied the two cases with separated Wilson line masses--even though this can occur only in special regions of moduli space. Under the assumption that the soft SUSY breaking masses be of TeV order--to assure that sparticle masses potentially be LHC accessible--we found that gauge coupling unification dictates that the Wilson line scales must be separated by less than/approximately an order of magnitude in either case. Additionally, we found that both cases lead to similar low energy phenomenology. Hence, for specificity, we carried out our analysis using the first of these symmetry breaking patterns; that is, the intermediate regime containing the ``left-right'' model. We refer the reader to~\cite{Ovrut:2012wg} for that analysis. Here, for concreteness, we simply show in Figure \ref{fig:MU.MI.MSUSY} the relationship of the $M_{U} \simeq M_{\chi_{B-L}}$ unification scale to that of the mass $M_{\chi_{3R}}$ of the the second Wilson line in the ``left-right'' model case. This is plotted  as a function of $M_{\susy}$ -- defined below in Eq. \eqref{eq:358}.

In this paper, we turn to the analysis of the {\it generic} region of moduli space where equation \eqref{s1}, that is, $M_{U} \simeq M_{\chi_{B-L}} \simeq M_{\chi_{3R}}$,  is satisfied, thereby {\it giving up exact gauge unification}. Be that as it may, to enable direct comparison of our new simultaneous Wilson line results with those from the split Wilson lines analyzed in~\cite{Ovrut:2015uea}, we continue to use the same notation for all quantities. In particular, it is important to use identical notation for the $B-L$ gauge coupling. Thus far in this paper, we have discussed the gauge parameter $g_{BL}$, which couples to the $\frac{I_{BL}}{2}$ generator. However, as was discussed in~\cite{Ovrut:2012wg}, this gauge coupling has to be properly normalized so as to unify with the other gauge parameters in the split Wilson line scenarios. The appropriate coupling was denoted $g_{BL}'$ and defined by
\beq
	g_{BL}' = \sqrt{\frac{2}{3}} g_{BL} \ .
\eeq	
Even though the four gauge couplings, including $g_{BL}' $, will not unify in the simultaneous Wilson line scenario in this paper, we will continue to use this parameter when appropriate. Note that $g_{BL}'$ couples to the $\sqrt{\frac{3}{8}} I_{BL}$ generator and will appear in the RGEs. For quantities of physical interest, such as physical masses, $g_{BL}$ will be used. 

To fully understand the evolution of this model from ``unification'' to the electroweak scale, it should be noted that there are four relevant mass scales of interest. All four are described in the following:\\

\noindent\textbf{$M_U$: the unification and the first and second Wilson lines mass scale.}

\begin{figure}
\centering
\includegraphics[scale=1.2]{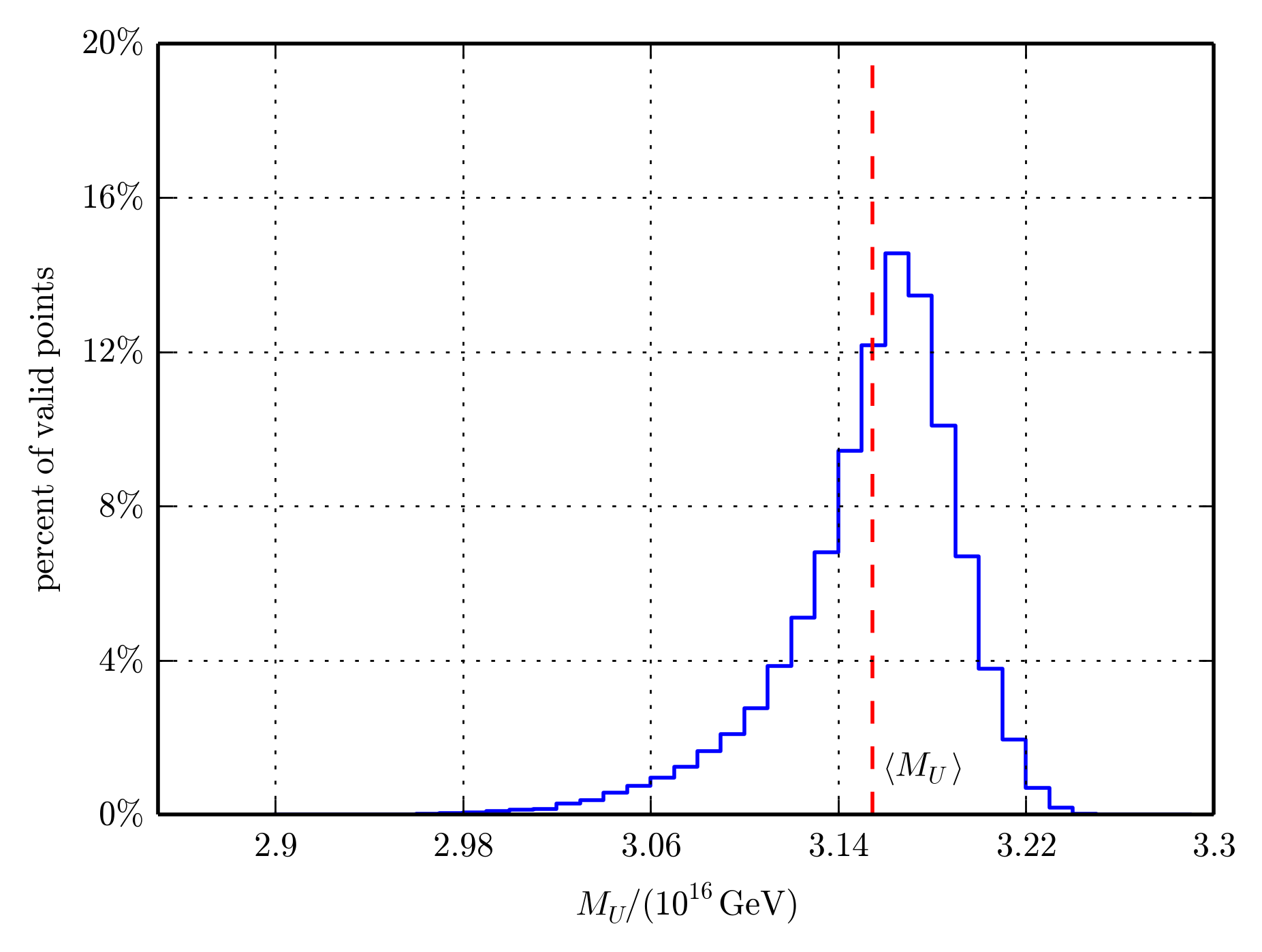}
\caption{A histogram of the unification scale for the 53,512 phenomenologically valid points in the split Wilson line ``left-right'' unification scheme. The average unification scale is $\left<M_U\right>=3.15\times10^{16}$ GeV.}
\label{fig:a}
\end{figure}

\begin{figure}
\centering
\includegraphics[scale=1.2]{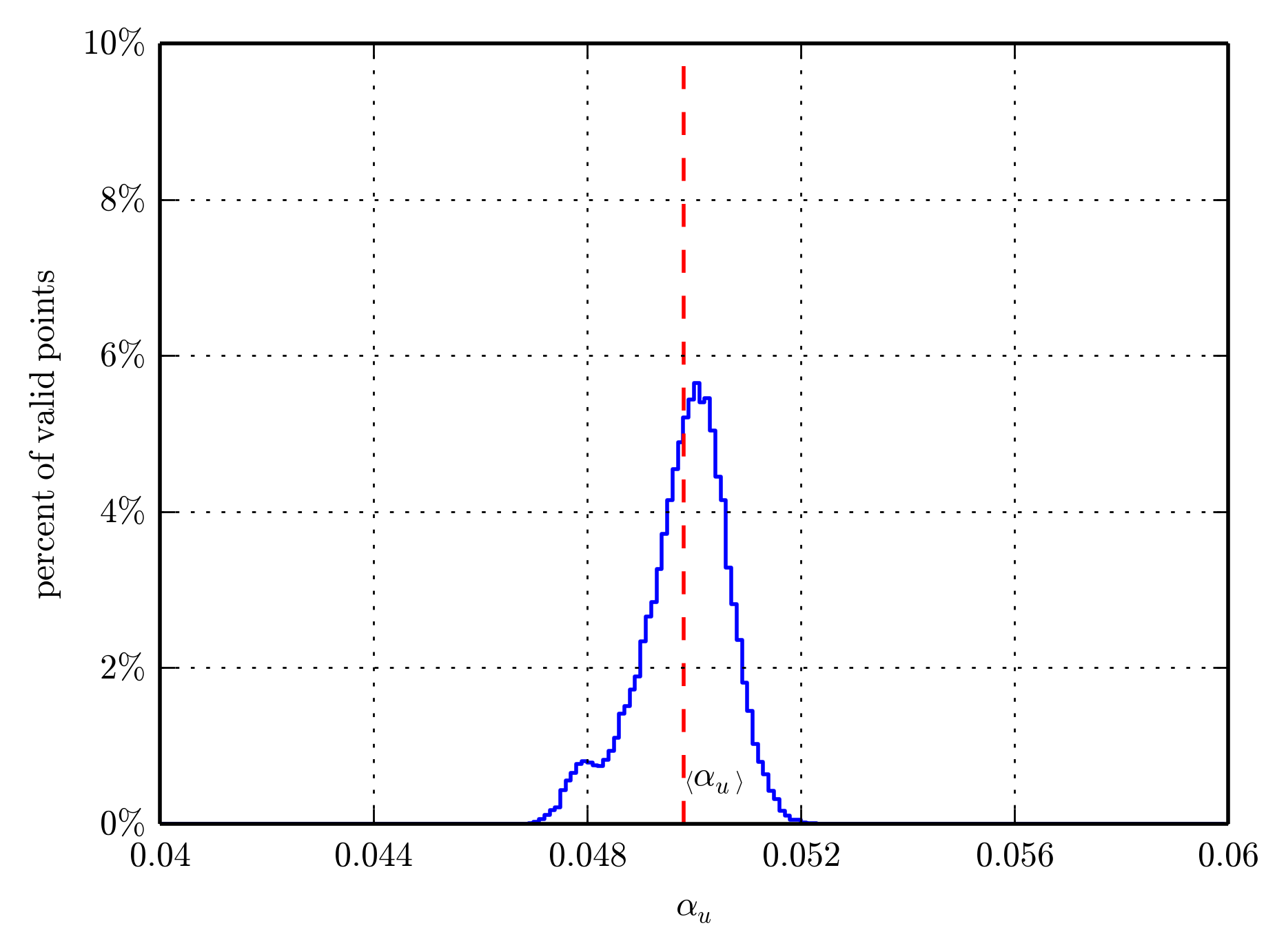}
\caption{A histogram of the unification scale for the 53,512 valid points in the split Wilson line ``left-right'' unification scheme. The average value of the unified gauge coupling is $\left<\alpha_u\right>=0.0498$.}
\label{fig:b}
\end{figure}

Since, as discussed above, exact unification of the four gauge couplings no longer occurs for simultaneous Wilson lines, it is essential to give a justification--and an explicit definition--of what we mean by the ``unification mass'' in the present context. In~\cite{Ovrut:2015uea}, every phenomenologically valid point in the space of randomly chosen initial soft supersymmetry breaking parameters  corresponds to an explicit unification mass $M_{U}$ and unified coupling $\alpha_{u}$. Both the unification scale and unification parameter vary for different valid points. The associated statistical histograms for these quantities are shown in Figures \ref{fig:a} and \ref{fig:b} respectively, along with their average values. These are found to be
\begin{equation}
\left< M_{U} \right> = 3.15 \times 10^{16}~\mbox{GeV} ~~, \quad \left< \alpha_{u} \right> = 0.0498 \ .
\label{s2}
\end{equation}
In this paper, we will refer to the average values $\left< M_{U} \right>$ and $\left< \alpha_{u} \right>$ as the ``unification'' mass and ``unified'' gauge coupling--and RG scale the gauge parameters between this scale and the electroweak scale. The values of the four diverse couplings $\alpha_{3},~ \alpha_{2}, ~\alpha_{3R}$ and $ \alpha_{BL}^{\prime}$ at $\left< M_{U} \right>$ will be determined for each statistical choice of soft supersymmetry breaking parameters. Henceforth, for specificity, we will always take this unification scale and both Wilson line masses to be strictly identical; that is
\begin{equation}
 \left<M_{U}\right> =  M_{\chi_{3R}} = M_{\chi_{B-L}} \ .  
 \label{hani1}
 \end{equation}
\\

\noindent\textbf{$M_{B-L}$: the mass at which the right-handed sneutrino VEV triggers $U(1)_{3R}\times U(1)_{B-L} \to U(1)_Y$ symmetry breaking.}

Physically, this corresponds to the mass of the neutral gauge boson $Z_R$ of the broken symmetry and, therefore, the scale of $Z_R$ decoupling. Specifically
	\beq
		M_{Z_R} = M_{B-L}.
		\label{eq:249}
	\eeq
Note that $M_{Z_R}$ itself depends on parameters evaluated at $M_{B-L}$. This results in a transcendental equation that can be solved using for $M_{B-L}$ using numerical methods. The boundary condition relating the hypercharge coupling to the gauge couplings of $U(1)_{3R}$ and $U(1)_{B-L}$ at this scale is nontrivial. It is given by
	\begin{equation}
		\label{eq:1.3R.BL}
		g_1 = \sqrt{\frac{5}{3}} g_{3R} \sin \theta_R = \sqrt{\frac{5}{2}} g_{BL}'\cos \theta_R \ ,
	\end{equation}
	where
	\beq
		\cos \theta_R = \frac{g_{3R}}{\sqrt{g_{3R}^2+\frac{3}{2} g_{BL}'^2}} \ .
	\label{home1}	
	\eeq
As with the $B-L$ gauge coupling, the hypercharge coupling has been rescaling to allow for unification in the split Wilson line scenarios. The rescaled hypercharge gauge coupling, $g_1$, is defined by
	\beq
		g_1 = \sqrt{\frac{5}{3}} g_Y \ .
	\label{home2}	
	\eeq
\\

\noindent\textbf{$M_\text{SUSY}$: the soft SUSY breaking scale.}

This is the scale at which all sparticles are integrated out, with the exception of the right-handed sneutrinos, which are associated with $B-L$ breaking and, therefore, are integrated out at the $B-L$ scale~\cite{Ovrut:2015uea}. While the sparticles do not all have the same mass, we use the scale of stop decoupling as a representative scale associated with all sparticles. That is,
\beq
	M_\text{SUSY} = \sqrt{m_{\tilde t_1}\ m_{\tilde t_2}}.
	\label{eq:358}
\eeq
The scale of stop decoupling is the best choice for the SUSY scale since the stops give the dominant radiative corrections to phenomenologically important quantities such as the electroweak scale and the Higgs mass. See, for example,~\cite{Gamberini:1989jw} for more details. Note that the physical stop masses depend on quantities evaluated at $M_\susy$. Therefore, this equation must be solved using iterative numerical methods for the correct value of $M_\susy$.\\

\noindent\textbf{$M_{\text{EW}}$: the electroweak scale.}

This is the well-known scale associated with the $Z$ and $W$ gauge bosons of the standard model (SM). We identify this scale with the mass of $Z$ boson, as is conventional. That is,
	\beq
		M_{\text{EW}} = M_Z.
	\eeq


\section{The Physical Regimes and the RG Scaling of the Supersymmetric Parameters}


Having defined the relevant mass scales, we turn to a brief discussion of RG evolution that occurs between them. The gauge coupling RGEs are
\beq
	\frac{d}{d t} \alpha_a^{-1} = -\frac{b_a}{2 \pi} \ ,
\eeq
where $a$ indexes the associated gauge groups. The slope factors $b_a$ are different in the different scaling regimes.
\begin{itemize}
	\item $\left<M_{U}\right> - {\rm max}(M_{\text{SUSY}}, M_{B-L})$: We refer to this regime as the ``$B-L$ MSSM regime'' because the particle content and gauge group are the $B-L$ MSSM. The $b_a$ factors are
	\begin{equation}
		b_3 = -3 ,\ b_2 = 1,\ b_{3R}= 7,\ b_{BL'}= 6 \ .
	\label{red1}
	\end{equation}
\end{itemize}
Note that the hierarchy between the SUSY and $B-L$ scales depends on the point chosen in the initial parameter space. The remaining two regimes depend on which of the following two cases occurs: $M_{B-L} > M_{\text{SUSY}}$--the ``right-side-up'' hierarchy--and $M_{\text{SUSY}} > M_{B-L}$--the ``upside-down'' hierarchy. \\

\begin{figure}
\centering
\includegraphics[scale=1.2]{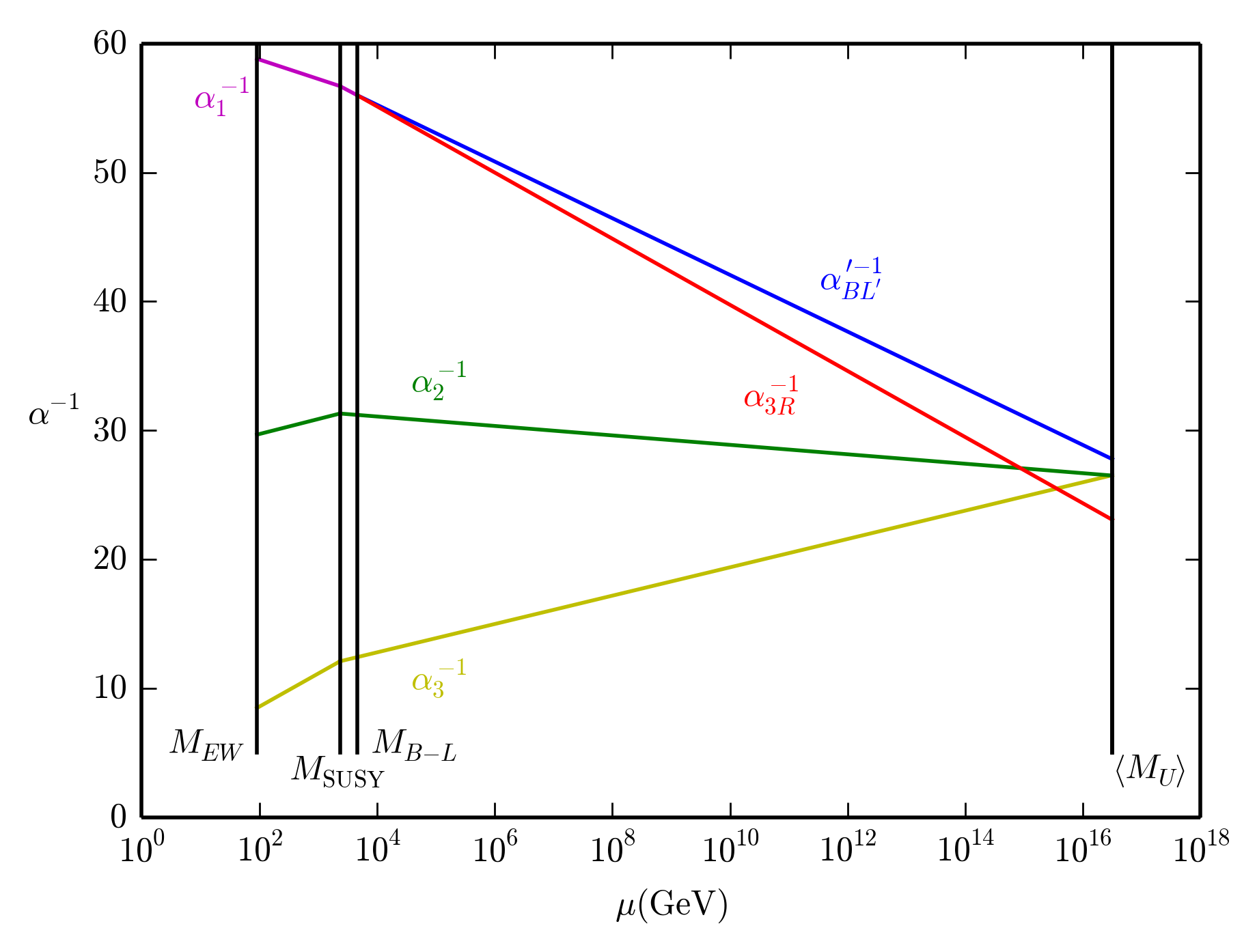}
\caption{Running gauge couplings for one of the valid points in our main scan, discussed below, with $M_\susy=2350$ GeV, $M_{B-L}=4670$ GeV and $\sin^{2}\theta_R = 0.6$. In this example, $\alpha_3(\left<M_U\right>)=0.0377$, $\alpha_2(\left<M_U\right>)=0.0377$, $\alpha_{3R}(\left<M_U\right>)=0.0433$, and $\alpha_{BL^\prime}(\left<M_U\right>)=0.0360$.}
\label{fig:c}
\end{figure}

\noindent \underline{right-side-up hierarchy}:
\begin{itemize}
	\item $M_{B-L} - M_{\text{SUSY}}$: In this regime, the gauge group and particle content is that of the MSSM plus two right-handed neutrino supermultiplets. The gauge couplings in this regime evolve with the slope factors
	\begin{equation}
		b_3 = -3,\ b_2 = 1,\ b_{1}=\frac{33}{5} \ .
	\end{equation}
We refer to this regime as the ``MSSM'' regime.
	\item $M_{\text{SUSY}} - M_{\text{EW}}$: In this regime, the sparticles are integrated out, leaving the SM with an additional two sterile neutrinos. It has the well-known slope factors
	\begin{equation}
		b_3 = -7,\ b_2 = -\frac{19}{6},\ b_{1}=\frac{41}{10} \ .
		\label{eq:644}
	\end{equation}
We refer to this regime as the ``SM'' regime.
\end{itemize}
\noindent \underline{upside-down hierarchy}:
\begin{itemize}
	\item $M_{\susy} - M_{B-L}$: In this regime, sparticles, with the exception of the third-generation right-handed sneutrino, are integrated out. But $B-L$ is still a good symmetry. This yields a non-SUSY $SU(3)_C \times SU(2)_L \times U(1)_{3R} \times U(1)_{B-L}$ model, which also includes three generations of right-handed sneutrinos--the third of which acts as the $B-L$ Higgs. The slope factors are
	\begin{equation}
		b_3 = -7,\ b_2 = \frac{19}{6},\ b_{3R}= \frac{53}{12}, \ b_{BL'} = \frac{33}{8}.
	\end{equation}
	\item $M_{B-L} - M_{\text{EW}}$: This regime is identical to the SM regime with slope factors given in Eq.~(\ref{eq:644}).
\end{itemize}
The boundary conditions imposed on the gauge couplings are that the three $\alpha_{i}$ coefficients of the SM take their experimental values at $M_Z$~\cite{PDG}:
\begin{equation}
	\label{eq:alpha.ew}
	\alpha_3(M_Z)=0.118,\ \alpha_2(M_Z)=0.0337,\ \alpha_1(M_Z)=0.0170 \ .
\end{equation}
These experimental values will then be scaled up through the various regimes:  $M_{EW} \rightarrow M_{SUSY}$, $M_{SUSY} \rightarrow M_{B-L}$
(\rm for the right-side-up hierarchy) or $M_{EW} \rightarrow M_{B-L}$, $M_{B-L} \rightarrow M_{SUSY}$ (\rm  for the upside-down hierarchy), followed by scaling through the 
 $B-L$ MSSM regime to $\left<M_{U}\right>$ using the beta functions listed above. The ``splitting'' of $\alpha_{1}$ to $\alpha_{3R}$ and $\alpha^{\prime}_{BL}$ at $M_{B-L}$ is achieved using the boundary conditions  \eqref{eq:1.3R.BL}, \eqref{home1}. In previous work~\cite{Ovrut:2012wg}, exact unification conveniently specified $\sin^2\theta_R\approx 0.6$. However, in the present scenario we are not requiring exact unification. Hence, this specificity is lost and $\sin^2\theta_R$ is a free parameter. We proceed by simply setting
 \begin{equation}
  \sin^2\theta_R=0.6
  \label{burt1}
  \end{equation}
in order to make the results of this paper more directly comparable to those of~\cite{Ovrut:2012wg}. An example of the running of the gauge couplings from the electroweak scale to $\left<M_{U}\right>$, as well as the values of the couplings $\alpha_{3}(\left<M_{U}\right>),~ \alpha_{2}(\left<M_{U}\right>), ~\alpha_{3R}(\left<M_{U}\right>)$ and $ \alpha_{BL}^{\prime}(\left<M_{U}\right>)$, is presented in Figure \ref{fig:c} using a phenomenologically acceptable point in the space of initial soft SUSY breaking parameters. 

 Similarly to the gauge parameters, the Yukawa couplings run differently under the RG through each of the above scaling regimes. Before discussing them, we must first decide which Yukawa couplings are relevant to our analysis. As discussed in \cite{{Ovrut:2015uea}}, we begin by inputting the experimentally determined Yukawa couplings derived from the fermion masses at the electroweak scale. For the purposes of this paper, the SM Yukawa couplings, which are three-by-three matrices in flavor space,  can all be approximated to be zero except for the three-three elements which give mass to the third-generation SM fermions. The experimentally determined initial conditions are
\beq
	y_t=0.955,\quad y_b=0.0174,\quad y_\tau=0.0102.
\label{home3}	
\eeq
For details on relating fermion masses to Yukawa couplings, see~\cite{Djouadi:2005gi}. We use lower case $y$ to denote Yukawa couplings in the non-SUSY regime. The one-loop RGEs for these Yukawa couplings were presented in Appendix A of \cite{{Ovrut:2015uea}}, to which we refer the reader. The Yukawa couplings in Eqn. \eqref{home3} can be evolved to the $\left<M_{U}\right>$ scale as follows. For the right-side-up scenario, the RGEs from $M_{EW} \rightarrow M_{SUSY}$ are given in Eqs.~(A5) -~(A7). There are non-trivial boundary conditions at the SUSY scale given by
\begin{eqnarray}
	y_{t}(M_\susy)&=&Y_{t}(M_\susy)\sin\beta\nonumber\\
	y_{b,\tau}(M_\susy)&=& Y_{b,\tau}(M_\susy)\cos\beta.
\label{home4}
\end{eqnarray}
From $M_{SUSY} \rightarrow M_{B-L}$, these parameters evolve as in (A12)~-~(A13). The boundary condition at the $B-L$ scale is trivial. Finally, from $M_{B-L} \rightarrow \left<M_{U}\right>$ the $B-L$ MSSM RGEs are given in Eqs.~(A14)~-~(A16). For the upside-down case, one evolves from $M_{EW} \rightarrow M_{B-L}$ using (A5)~-~(A7), as previously. However, between  $M_{B-L} \rightarrow M_{SUSY}$ the RGEs are given in Eqs.~(A8)~ -~(A10). Finally, above the SUSY scale one uses the $B-L$ MSSM equations given in (A14)-(A16).


\section{The Soft Supersymmetry Breaking Parameters}


The remaining parameters of the $B-L$ MSSM are the massive coefficients appearing in Eqn.~\eqref{home5} which are responsible for softly breaking supersymmetry. Their RGEs in each physical regime were presented in detail in \cite{{Ovrut:2015uea}} and won't be discussed in this paper. Here, we simply note that flavor and CP-violation experimental results place well-known limits on these quantities. Generically, the implication of these constraints are, approximately, as follows:
\begin{itemize}
	\item Soft sfermion mass matrices are diagonal.
	\item The first two generations of squarks are degenerate in mass.
	\item The trilinear $a$-terms are diagonal.
	\item The gaugino masses and trilinear $a$-terms are real.
\end{itemize}
It is typically assumed that the soft trilinear $a$-terms are proportional to the Yukawa couplings. That is, $a = Y A$ for each fermions species. Each $A$ is real and associated with the SUSY scale. Each $Y$ factor is a dimensionless matrix in family space. This condition effectively makes all non-third-generation trilinear terms insignificant. The above constraints are summarized as
\begin{align}
\begin{split}
	& m_{\tilde q}^2 = \text{diag}\left(m^{2}_{\tilde q_1},m^{2}_{\tilde q_1}, m^{2}_{\tilde q_3} \right)~~,~~
	\tilde q = \tilde Q, \, {\tilde u}^{c}, \,  {\tilde d}^{c} \ ,
	\\
	&  m_{\tilde \ell}^2 = \text{diag}\left(m^{2}_{\tilde \ell_1},m^{2}_{\tilde \ell_2}, m^{2}_{\tilde \ell_3} \right)~,~~
	 \tilde \ell = \tilde L, \, {\tilde e}^{c} \ , \tilde \nu^c \ ,
	\\
	& a_f = Y_fA_f ~~~~,~~ \ f = t,\, b, \, \tau \ .
\end{split}
\end{align}
These constraints can be implemented at the scale $\left<M_{U}\right>$, since RG evolution to the SUSY scale will not spoil these relations. Note that we do not assume that the first and second generation slepton masses are degenerate, unlike the squark masses, since this is not required by experiments. The degeneracy or non-degeneracy of the first and second generation sleptons will not, however, greatly effect the results of this paper. 

We now turn to the input values for the SUSY breaking parameters. Unlike the cases of the gauge and Yukawa couplings, these soft SUSY breaking parameters are not experimentally determined. In \cite{{Ovrut:2015uea}}, we introduced a novel way to analyze the initial parameter space of a SUSY model. We will follow the same approach in the present analysis of simultaneous Wilson lines. Specifically, we run a statistical scan of input parameters at the scale $\left<M_U\right>$. The randomly generated input parameters are then RG evolved to the SUSY scale. We conduct an analysis of which of these high-scale initial conditions lead to realistic physics. Although the soft SUSY breaking Lagrangian contains over 100 dimensionful parameters, the phenomenologically motivated assumptions discussed briefly above only allow significant values for 24 of them. These, along with $\tan \beta$ and the sign of certain parameters, are presented in the first column of Table~\ref{tbl:scan}.

The high-scale initial values of the 24 relevant dimensionful SUSY breaking parameters are determined as follows. We make the assumption that there is only one overall scale associated with SUSY breaking, requiring that these parameters be separated by no more than an order of magnitude, or so, from each other. To quantify this, we demand that any dimension one soft SUSY breaking parameter be chosen at random within the range
\beq
	(\frac{M}{f}, Mf) \ ,
\eeq
where $M$ is the overall scale of SUSY breaking and $f$ is a dimensionless number satisfying $1\leq f  \lesssim 10$. We will further insist that any such parameter be evenly scattered around $M$; that is, that $M$ be the average of the randomly generated values.  In \cite{{Ovrut:2015uea}}, we found that in the case of split Wilson line masses, the maximal number of phenomenologically acceptable ``valid'' initial points were obtained by statistically scattering within the interval defined by
\begin{eqnarray}
	M=2700 \text{ GeV} ,\ f=3.3 \ .
\label{cat1}
\end{eqnarray}
To allow direct comparison of the results of this paper to those of \cite{{Ovrut:2015uea}}, we will continue to use these values in the present context. This is shown in the second column of Table~\ref{tbl:scan}, along with the scattering interval associated with tan${\beta}$ and the allowed signs of various parameters.

\begin{table}[htdp]
\begin{center}
\begin{tabular}{|c|c|c|}
	\hline
	Parameter & Range 
	\\
	\hline
	\hline
	\quad $m_{\tilde q_1} = m_{\tilde q_2}, \ m_{\tilde q_3}: \quad \tilde q = \tilde Q, \tilde u^c, \tilde d^c$ \quad & (820, 8900) GeV 
	\\
	$m_{\tilde \ell_1}, m_{\tilde \ell_2}, \ m_{\tilde \ell_3}: \quad \tilde \ell = \tilde L, \tilde e^c, \tilde \nu^c$
		& \ (820, 8900) GeV 
	\\
	$m_{H_u}, m_{H_d}$ & (820, 8900) GeV
	\\
	$\left|A_f\right|: \quad f = t,b, \tau$ & (820, 8900) GeV
	\\
	$\left|M_a\right|: \quad a = 3R, BL^\prime, 2, 3$ & (820, 8900) GeV
	\\
	$\tan \beta$ & (1.2, 65) 
	\\
	Sign of $\mu, a_f, M_a: \quad f=t,b,\tau \quad a=3R, BL^\prime, 2, 3$ & [-,+] 
	\\
	\hline
\end{tabular}
\end{center}
\caption{The parameters and their ranges scanned in this study. The ranges for the soft SUSY breaking parameters are taken to be those of \cite{{Ovrut:2015uea}}.}
\label{tbl:scan}
\end{table}%


\section{The Parameter Scan and Results}


The technical details of our statistical scan over the interval of soft supersymmetry breaking parameters, the complete set of all RG equations, the evolution of all parameters under the RGEs and a discussion of the sparticle and the Higgs masses were presented in detail in both the text and Appendices of \cite{{Ovrut:2015uea}}. We will not repeat them here and refer the reader to that paper. In this section, we will simply apply these methods to the more ``natural'' case of simultaneous Wilson lines satisfying Eqn.~\eqref{hani1}. As we did in  \cite{{Ovrut:2015uea}}, we will perform a scan over 10 million random initial points, searching for those ``valid'' points that satisfy all present experimental lower bounds on the masses of the different types of SUSY particles and the $B-L$ gauge boson. These lower bounds are presented in Table~\ref{tbl:mass.bounds}.
\begin{table}[htdp]
\begin{center}
\begin{tabular}{|c|c|}
\hline
Particle(s) & Lower Bound
\\
\hline
\hline
Left-handed sneutrinos & 45.6 GeV
\\
$\quad$Charginos, sleptons $\quad$& 100 GeV
\\
Squarks, except for stop or sbottom LSP's & 1000 GeV
\\
Stop LSP (admixture)							& 450 GeV
\\
Stop LSP (right-handed)							& 400 GeV
\\
Sbottom LSP							& 500 GeV
\\
Gluino & 1300 GeV
\\
$Z_R$ & 2500 GeV
\\
\hline
\end{tabular}
\end{center}
\caption{The different types of SUSY particles and the lower bounds implemented in this paper.}
\label{tbl:mass.bounds}
\end{table}%
In addition, we will impose the requirement that the Higgs mass be within the $2\sigma$ allowed range from the value measured at the ATLAS experiment at the LHC~\cite{Aad:2014aba,Chatrchyan:2013lba}:
\begin{eqnarray}
	m_{h^0}=125.36\pm 0.82\mbox{ GeV}.
\end{eqnarray}
Since the initial soft SUSY breaking parameter space is 24-dimensional, graphically displaying the results is, in principle, very difficult. However, as was discussed in both the text and Appendices of \cite{{Ovrut:2015uea}}, much of the scaling behavior of the parameters is controlled by the two $S$-terms, $S_{BL^\prime}$ and $S_{3R}$, defined by
\begin{eqnarray}
	\label{eq:S.BLA}
	&S&_{BL^\prime}=\Tr(2m_{\tilde Q}^2-m_{\tilde u^c}^2-m_{\tilde d^c}^2-2m_{\tilde L}^2+m_{\tilde \nu^c}^2+m_{\tilde e^c}^2) \ ,\\
	\label{eq:S.RA}
	&S&_{3R}=m_{H_u}^2-m_{H_d}^2+\Tr\left(-\frac{3}{2}m_{\tilde u^c}^2+\frac{3}{2}m_{\tilde d^c}^2-\frac{1}{2} m_{\tilde \nu^c}^2+\frac{1}{2} 	m_{\tilde e^c}^2\right) \ ,
\end{eqnarray} 
where ``Tr'' implies a sum over the three families. It follows that our results can be reasonably displayed in the two-dimensional  $S_{BL^\prime}(\left<M_{U}\right>)$ - $S_{3R}(\left<M_{U}\right>)$ plane. 

\begin{figure}
	\centering
	\includegraphics[scale=1.2]{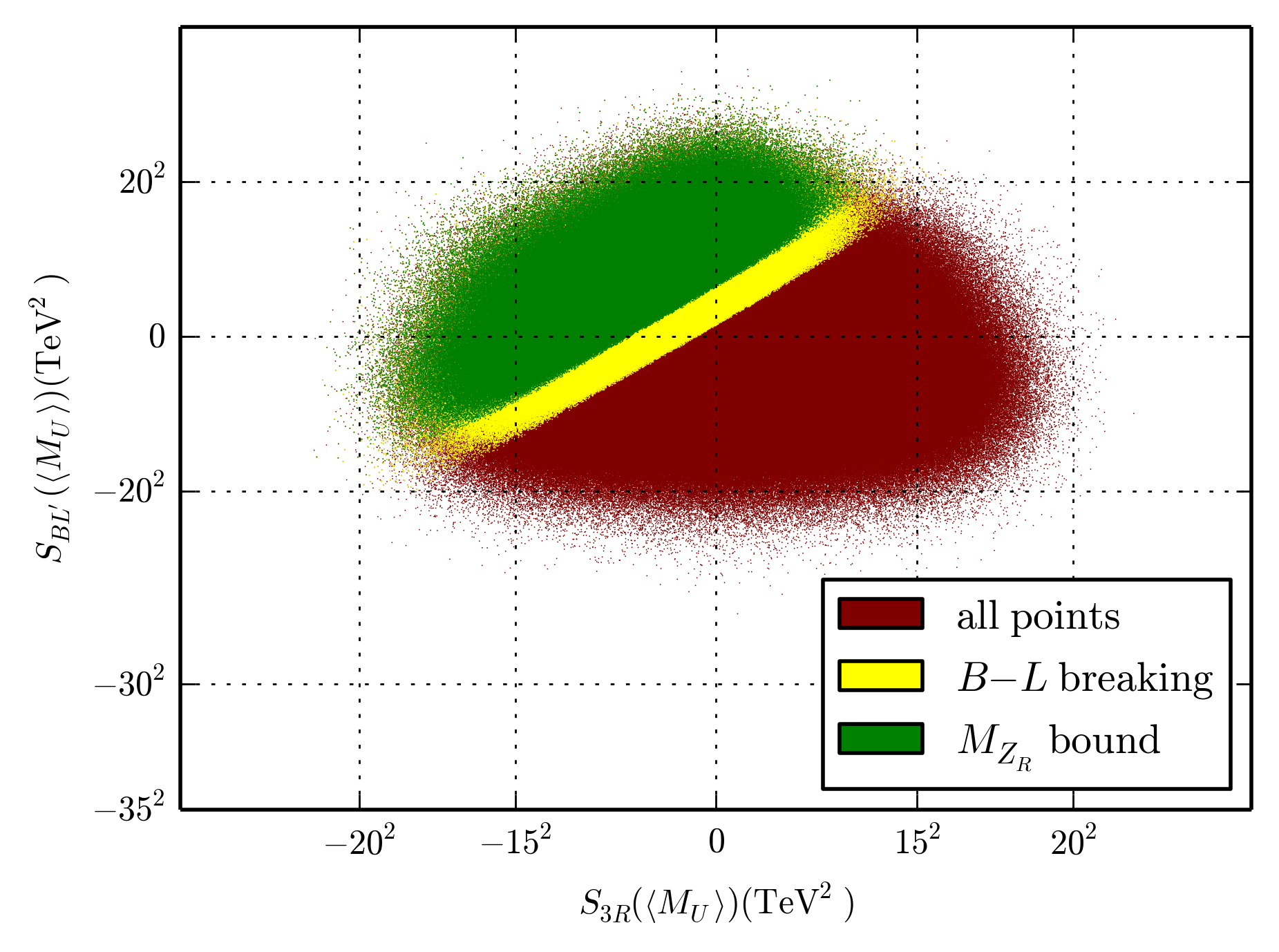}
	\caption{\small Points from the main scan in the $S_{BL^\prime}(\left<M_{U}\right>)$ - $S_{3R}(\left<M_{U}\right>)$ plane. Red indicates no $B-L$ breaking, in the yellow  region $B-L$ is broken but the $Z_R$ mass is not above its 2.5 TeV  lower bound, while green points have both $B-L$ breaking and $M_{Z_R}$ above this bound. The figure expresses the fact that, despite there being 24 parameters at the UV scale scanned in our work, $B-L$ physics is essentially dependent on only two combinations of them--the two $S$-terms. 
Note that the green points obscure some yellow and red points behind them. Similarly the yellow points obscure some red points.}
	\label{fig:1204}
\end{figure}

We begin by presenting in Fig.~\ref{fig:1204} all 10 million initial points in the $S_{BL^\prime}(\left<M_{U}\right>)$ - $S_{3R}(\left<M_{U}\right>)$ plane in order to explore, sequentially, which points satisfy the first two fundamental checks that we require; that is, $B-L$ breaking and the experimental $Z_{R}$ mass lower bound. Points that do not break $B-L$ are shown in red, points that satisfy $B-L$ breaking but not the $Z_R$ mass bound are in yellow, and points that break $B-L$ symmetry and satisfy the $Z_R$ mass bound are shown in green. We find that out of the 10 million initial points, 
\begin{itemize}
\item 1,629,001 --the green and yellow points-- break $B-L$ symmetry.
\item  697,886 --the green points--break $B-L$ with $M_{Z_{R} }>$ 2.5 TeV. 
\end{itemize}
\begin{figure}[!htbp]
        \centering
        \includegraphics[scale=1.2]{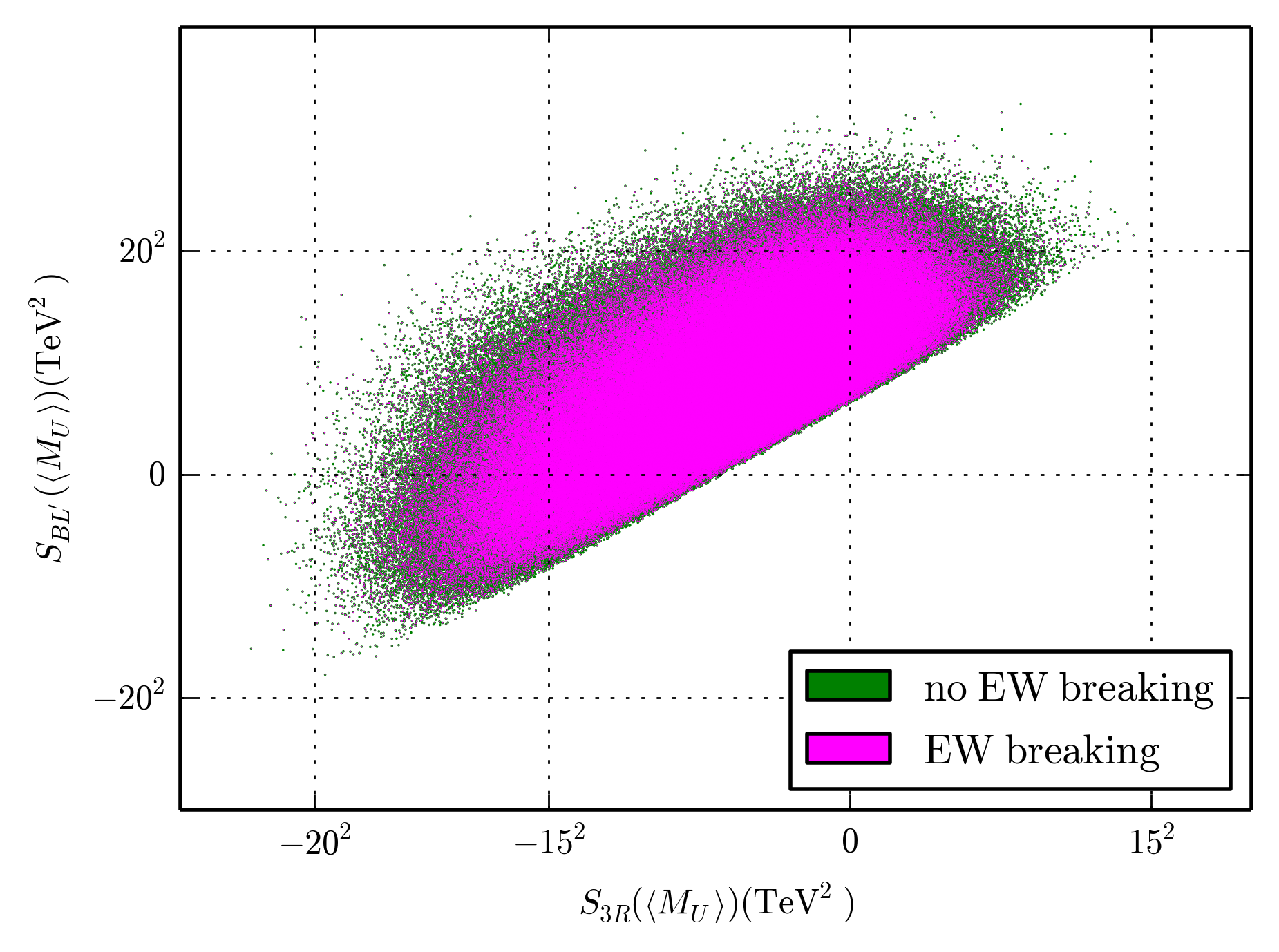}
	\caption{\small A plot encompassing the green region in Fig~\ref{fig:1204}. The green points in this plot correspond to those which appropriately break $B-L$ symmetry, but which do not break electroweak symmetry. However, the purple points, in addition to breaking $B-L$ symmetry with an appropriate $Z_{R}$ mass, also break EW symmetry. Note that a small density of green points that do not break EW symmetry are obscured by the purple points.}
        \label{fig:1205}
\end{figure}
\begin{figure}[!htbp]
	\centering
	\includegraphics[scale=1.2]{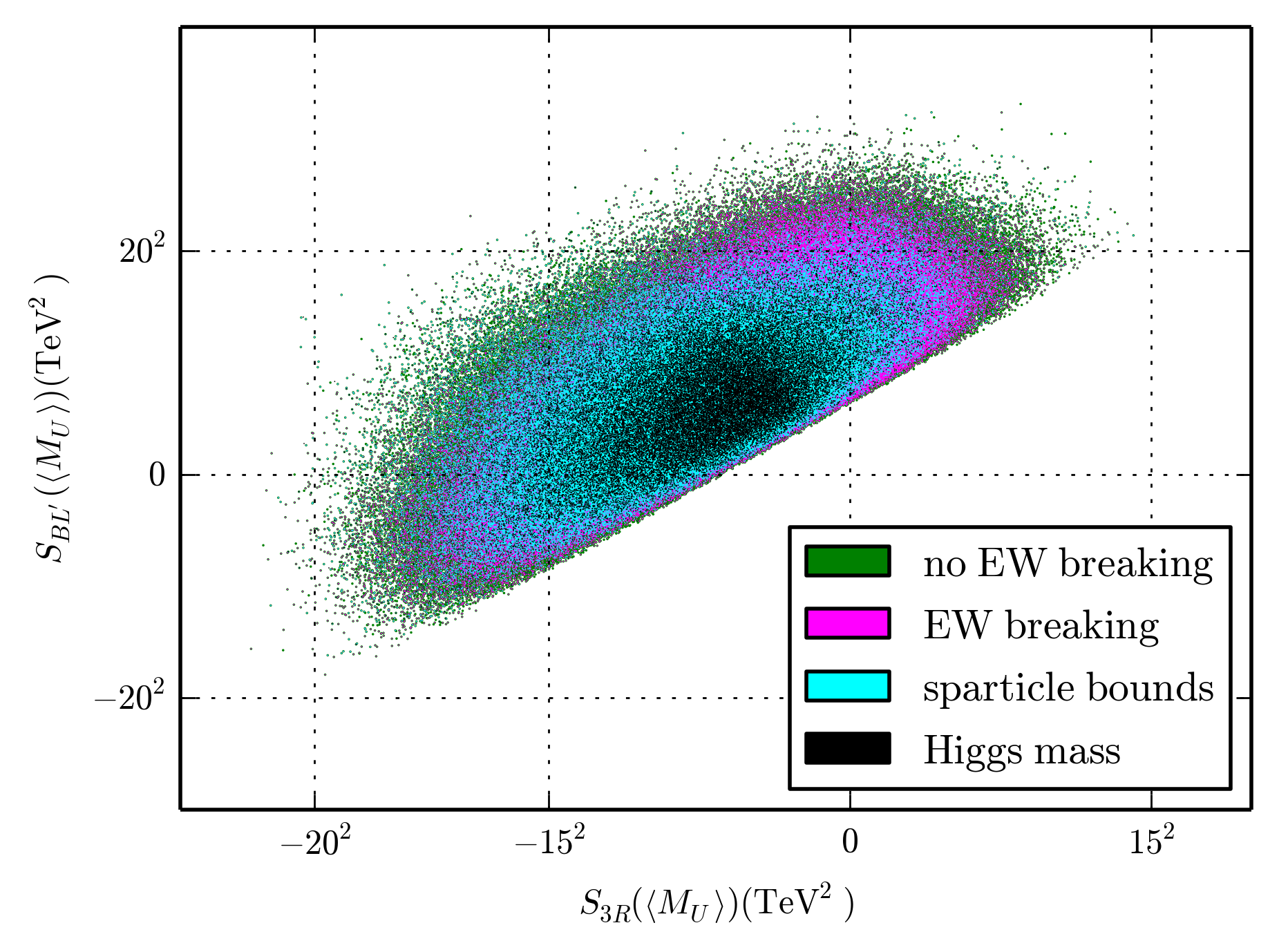}
	\caption{\small A plot of the ``valid'' points in our main scan. The green and purple points correspond to the green and purple points in Fig~\ref{fig:1205}. The cyan points additionally satisfy all sparticle mass lower bounds. The black points are fully valid. That means that, in addition to satisfying all previous checks, they reproduce the correct Higgs mass within the stated tolerance. The distribution of points indicates that while $B-L$ breaking prefers large $S$-terms, sfermion mass constraints prefer them to be not too large. Again, the cyan and black points may obscure a low density of other points not satisfying their constraint.}
	\label{fig:1148}
\end{figure}
This plot shows that $B-L$ breaking consistent with present experiments is a robust phenomena. Furthermore, it shows the strong dependence of $B-L$ breaking and the $Z_R$ mass on the values of the $S$-terms. There is a line in the $S_{BL^\prime}$ - $S_{3R}$ plane--between the yellow and red regions--below which $B-L$ breaking is not possible. Note that this includes the origin, which corresponds to vanishing $S$-terms and, hence, universal soft masses. This shows that at least a small splitting from sparticle universality is required for $B-L$ breaking. Another line exists--between the green and yellow regions--below which $Z_R$ is always lighter than its experimental lower bound.

Proceeding sequentially, we present in Fig.~\ref{fig:1205} the initial points in the $S_{BL^\prime}(\left<M_{U}\right>)$ - $S_{3R}(\left<M_{U}\right>)$ plane that, in addition to breaking $B-L$ with a $Z_{R}$ mass above the experimental bound, also break EW symmetry. The entire colored region encompasses the green points shown in Fig.~\ref{fig:1204}. Those points that also break EW symmetry are displayed in purple. This plot indicates that most of the points that break $B-L$ with a $Z_{R}$ mass above the experimental bound, also break EW symmetry. Note that a small density of green points that do not break EW symmetry are obscured by the purple points. Specifically, we find that out of the 697,886 green points that break $B-L$ with $M_{Z_{R} }>$ 2.5 TeV,
\begin{itemize}
\item 485,952 -- the purple points-- also break EW symmetry.
\end{itemize}

In Fig.~\ref{fig:1148}, we reproduce Fig.~\ref{fig:1205} but now, in addition, sequentially indicate the points that are consistent with the remaining checks--that is, all lower bounds on sparticles masses satisfied and, finally, that they reproduce the Higgs mass within the experimental uncertainty. Points that appropriately break $B-L$ symmetry but do not satisfy electroweak symmetry breaking are still shown in green. Points that, additionally, do break electroweak symmetry are again shown in purple. Such points that also satisfy all lower bounds on sparticles masses, but do not match the known Higgs mass, are now indicated in cyan. Finally, points that satisfy all checks, including the correct Higgs mass, are shown in black. These are the ``valid'' points. The density of black points indicate that there is a surprisingly high number of initial parameters that satisfy all present low energy experimental constraints. Specifically, we find that out of the 485,952 purple points that appropriately break $B-L$ symmetry as well as EW symmetry,
\begin{itemize}
\item 228,278 --the cyan points-- also satisfy all sparticle lower mass bounds.
\item 44,884 --the black points-- satisfy all sparticle lower mass bounds and also give the measured value of the Higgs mass.
\end{itemize} 
The distribution of black points can be explained from the fact that, while $B-L$ breaking favors non-zero $S$-terms, very large $S$-terms can effect the RGE evolution of sfermion masses adversely. Since the effect of the $S$-terms depends on the charge of the sfermion in question, some sfermions will become quite heavy while others light or tachyonic. Therefore, in general, the valid points in our scan are a compromise between large $S$-terms, needed for a $Z_R$ mass above its lower bound, and small $S$-terms needed to keep the sfermion RGEs under control.\\

\noindent {\bf The LSP Spectrum:}
 
\begin{figure}
	\centering
	\includegraphics[scale=1.2]{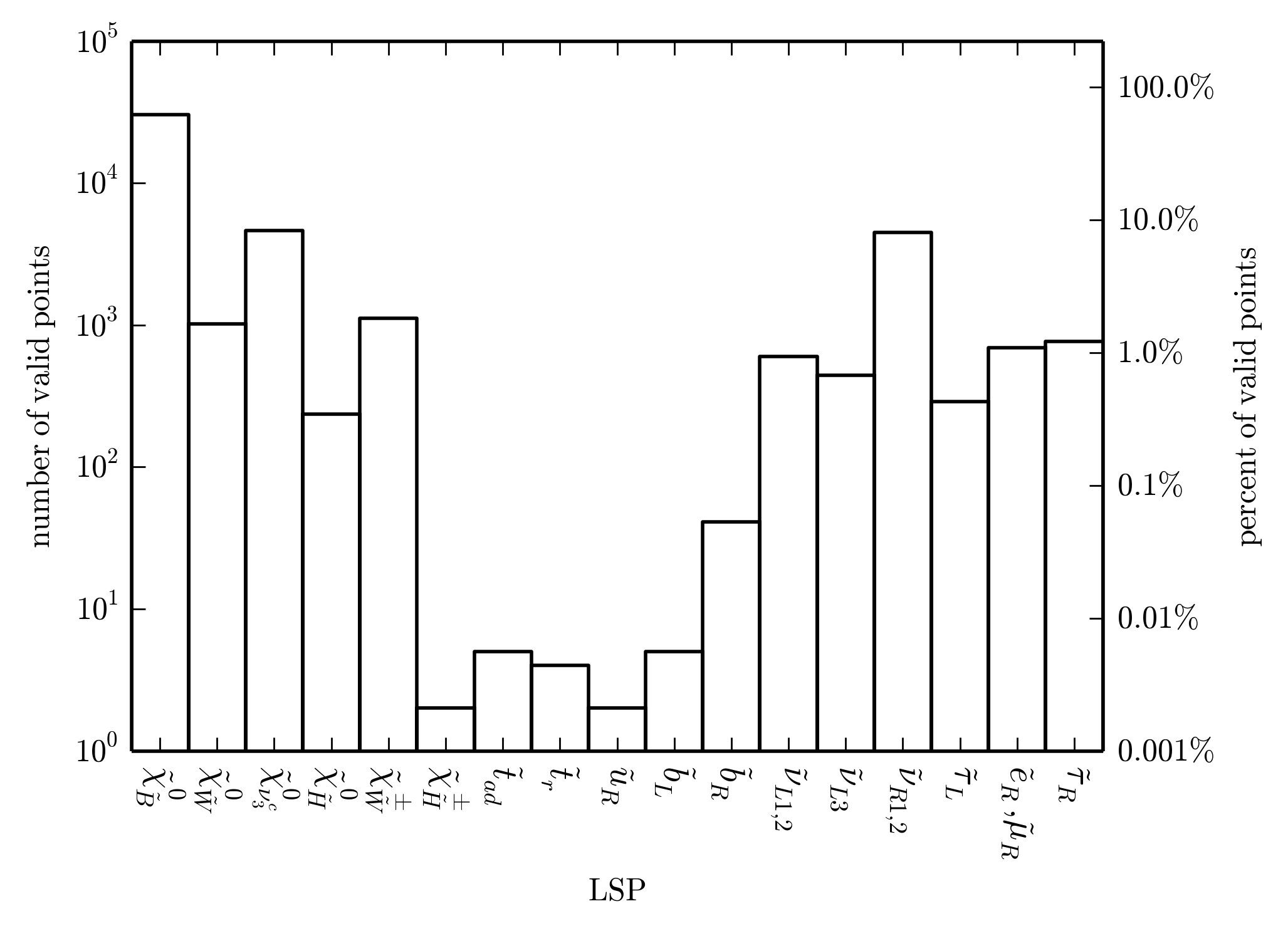}
	\caption{\small A histogram of the LSP's in the main scan showing the percentage of valid points with a given LSP. Sparticles which did not appear as LSP's are omitted. The y-axis has a log scale. The dominant contribution comes from the lightest neutralino, as one might expect. The notation for the various states, as well as their most likely decay products, are given in Table~\ref{tbl:LSP}. Note that we have combined left-handed first and second generation sneutrinos into one bin, and that each generation makes up about 50\% of the LSP's. The same is true for the first and second generation right-handed sleptons and sneutrinos.
}
	\label{fig:1039}
\end{figure}

An important property of the initial SUSY parameter space in determining low-energy phenomenology is the identity of the LSP. Recall that when R-parity is violated, no restrictions exist on the identity of the LSP; for example, it can carry color or electric charge. Our main scan provides an excellent opportunity to examine the possible LSP's and the probability of their occurrence . To this end, a histogram of possible LSP's is presented in Fig.~\ref{fig:1039}--with the possible LSP's indicated along the horizontal axis, and $\log_{10}$ of the number of valid points with a given LSP on the vertical axis. The notation here is a bit condensed, but is specified in more detail in Table~\ref{tbl:LSP}. The notation is devised to highlight the phenomenology of the different LSP's, specifically their decays\footnote {Recall that when R-parity is violated, as it is in this paper, the LSP can decay to lighter non-supersymmetric states.}, which are also presented in Table~\ref{tbl:LSP}.

\begin{table}[htdp]
\begin{center}
\begin{tabular}{|c|c|c|}
\hline
\ Symbol \ & Description & Decay
\\
\hline
\hline
$\tilde \chi^0_{\tilde B}$ & A bino-like neutralino, mostly rino ($\tilde W_R$) or mostly blino ($\tilde B'$).
& \multirow{4}{*}{$\ell^\pm W^\mp$, $\nu Z$, $\nu h$}
\\
$\tilde \chi^0_{\tilde W}$ & Mostly wino neutralino. &
\\
$\tilde \chi_{\nu^c}$ & Mostly third-generation right-handed neutrino. &
\\
$\tilde \chi^0_{\tilde H}$ & Mostly Higgsino neutralino. &
\\
\hline
$\tilde \chi^{\pm}_{\tilde W}$ & Mostly wino charginos. & \multirow{2}{*}{$\nu W^\pm$, $\ell^{\pm} Z$, $\ell^{\pm} h$}
\\
$\tilde \chi^{\pm}_{\tilde H}$ & Mostly Higgsino charginos. & 
\\
\hline
$\tilde g$ & Gluino. & $t \bar t \nu$, $t \bar b \ell^-$
\\
\hline
 $\tilde t_{ad}$ & Left- and right-handed stop admixture. & $\ell^+  b$
\\
\hline
$\tilde t_{r}$ & Mostly right-handed stop (over 99\%). & $t \nu$, $\tau^+ b$
\\
\hline
$\tilde q_R$ & Right-handed first and second generation squarks. & $\ell^+j$, $\nu j$
\\
\hline
$\tilde b_L$ & Mostly left-handed sbottom. & $b \nu$
\\
\hline
$\tilde b_R$ & Mostly right-handed sbottom. & $b \nu$, $\ell^- t$
\\
\hline
\multirow{2}{*}{$\tilde \nu_{L_{1,2}}$} & First and second generation left-handed sneutrinos.
				& \multirow{3}{*}{\parbox[t]{3cm}{$b \bar b$, $W^+ W^-$, $ZZ$, \\ $t \bar t$, $\ell'^+ \ell^-$, $hh$, $\nu \nu$}}
\\
								&	LSP's are split evenly among these two generations. &
\\
$\tilde \nu_{L_3}$ & Third generation left-handed sneutrino. & 
\\
\hline
$\tilde \nu_{R_{1,2}}$ & First and second generation right-handed sneutrinos. & $\nu \nu$
\\
\hline
\multirow{2}{*}{$\tilde \tau_L$} &  \multirow{2}{*}{Third generation left-handed stau.} & $t \bar b$, $W^- h$,
\\
			&		& $e \nu$, $\mu \nu$, $\tau \nu$
\\
\hline
\multirow{2}{*}{$\tilde e_R, \mu_R$} &  First and second generation right-handed sleptons. & \multirow{2}{*}{$e \nu$, $\mu \nu$}
\\
	& LSP's are split evenly between these two generations.	&
\\
\hline
$\tilde \tau_R$ &  Third generation right-handed stau. & $t \bar b$, $e \nu$, $\mu \nu$, $\tau \nu$
\\
\hline
\end{tabular}
\end{center}
\caption{The notation used for the states in Fig.~\ref{fig:1039} and their probable decays. More decays are possible in certain situations depending on what is kinematically possible and the parameter space. Gluino decays are especially dependent on the NLSP, here assumed to be a neutralino. Here, the word ``mostly'' means it is the greatest contribution to the state. The symbol $\ell$ represents any generation of charged leptons. The left-handed sneutrino decay into $\ell'^+ \ell^-$ indicates a lepton flavor violating decay--that is, $\ell'^+$ and $\ell^-$ do not have the same flavor.  Note that $j$ is a jet--indicating a light quark.}
\label{tbl:LSP}
\end{table}
The most common LSP in our main scan is the lightest neutralino, $\tilde \chi_1^0$. However, not all $\tilde \chi_1^0$ states are created equal. LHC production modes for the lightest neutralino depend significantly on the composition of the neutralino--a bino LSP cannot be directly produced at the LHC, but the other neutralino LSP's can. This is the basis we use for the division of these states. The state $\tilde \chi_{\tilde B}^0$ designates a mostly rino or mostly blino neutralino, $\tilde \chi_{\tilde W}^0$ a mostly wino neutralino and $\tilde \chi_{\tilde H}^0$ a mostly Higgsino neutralino. Here, the subscript mostly indicates the greatest contribution to that state. As an unrealistic example, if $\tilde \chi_1^0$ is 34\% wino, 33\% bino and 33\% Higgsino, it is still labeled $\tilde \chi_{\tilde W}^0$. The chargino LSP's are similarly separated into wino-like and Higgsino-like charginos, and the stops and sbottom divisions are as in our earlier papers, references~\cite{Marshall:2014kea,Marshall:2014cwa}. Note that this notation for the stops, $\tilde t_{ad}$ and $\tilde t_r$, are only used to describe stop LSP's. For non-LSP stops, we use the conventional notation $\tilde t_1$ and $\tilde t_2$.

To make Fig.~\ref{fig:1039} more readable, we have made an effort to combine bins that have similar characteristics. The first and second generation left-handed sneutrinos are combined into one bin, where about 50\% of the LSP's are first generation sneutrinos. The same holds true for the first and second generation right-handed sleptons, while the first generation right-handed sneutrino is always chosen to be lighter than the second generation right-handed sneutrino. This similarity between the first and second generation sleptons is expected, since their corresponding Yukawa couplings are not large enough to distinguish them through the RG evolution. For both sleptons and squarks, more LSP's exist for the third-generation--as expected from the effects of the third-generation Yukawa couplings, which tend to decrease sfermion masses in the RGE evolution.

The myriad of possible LSP's leads to a rich collider phenomenology. This phenomenology is not the main focus of this paper, but it is worthwhile to briefly review it here. In models where R-parity violation is parameterized by bilinear R-parity breaking, such as the $B-L$ MSSM, SUSY particles are still pair produced and cascade decay to the LSP. At this point, the bilinear R-parity violating terms allow the LSP to decay. While only a few studies have been done on the phenomenology of the minimal $B-L$ MSSM~\cite{FileviezPerez:2012mj, Perez:2013kla,Marshall:2014kea,Marshall:2014cwa}, there have been several works on the phenomenology of explicit bilinear R-parity violation, which has some similarities to this model. See~\cite{Porod:2000hv, Hirsch:2003fe, Graham:2012th, Graham:2014vya} for general discussions. Table~\ref{tbl:LSP} provides some basic information on the most probable decay modes of each of the possible LSP's. Note that $\ell$ signifies a charged lepton of any generation and $j$ a jet--implying a light quark. Some interesting aspects of Table~\ref{tbl:LSP} were discussed in  \cite{{Ovrut:2015uea}}.\\

\noindent {\bf The Non-LSP Spectrum:}

To get a sense of the non-LSP spectrum, we produce histograms of the masses of the sparticles associated with the valid points in the main scan. In the following histograms, there will be quite a few pairs of fields that will be highly degenerate; these will be represented by only one curve. This includes $SU(2)_L$ sfermion partners, which are only split by small electroweak terms. First generation squarks are also degenerate with second generation squarks with the same isospin, due to phenomenological constraints. A consequence of this is that all first and second generation left-handed squarks are highly degenerate.
\begin{figure}
\centering
	\includegraphics[scale=0.6]{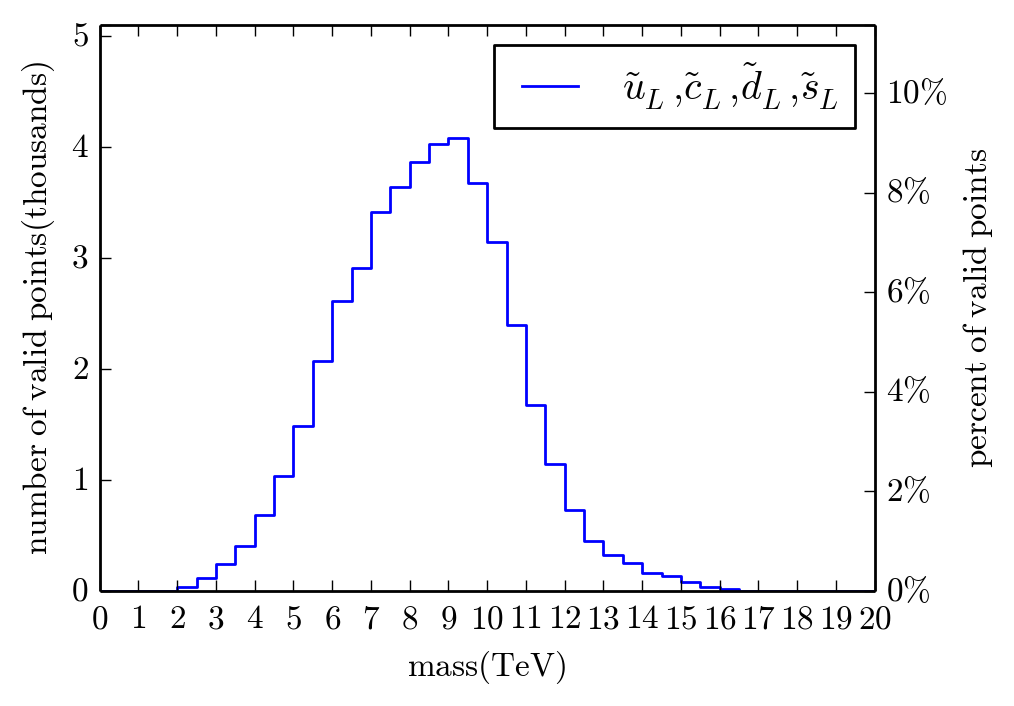}
	\includegraphics[scale=0.6]{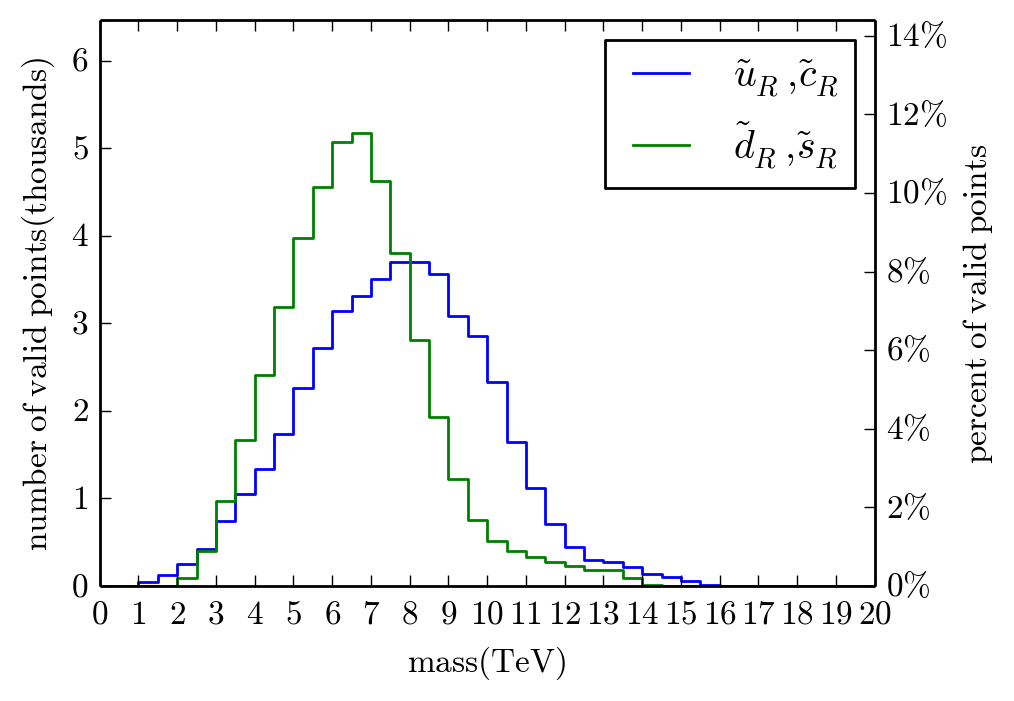}
	\includegraphics[scale=0.6]{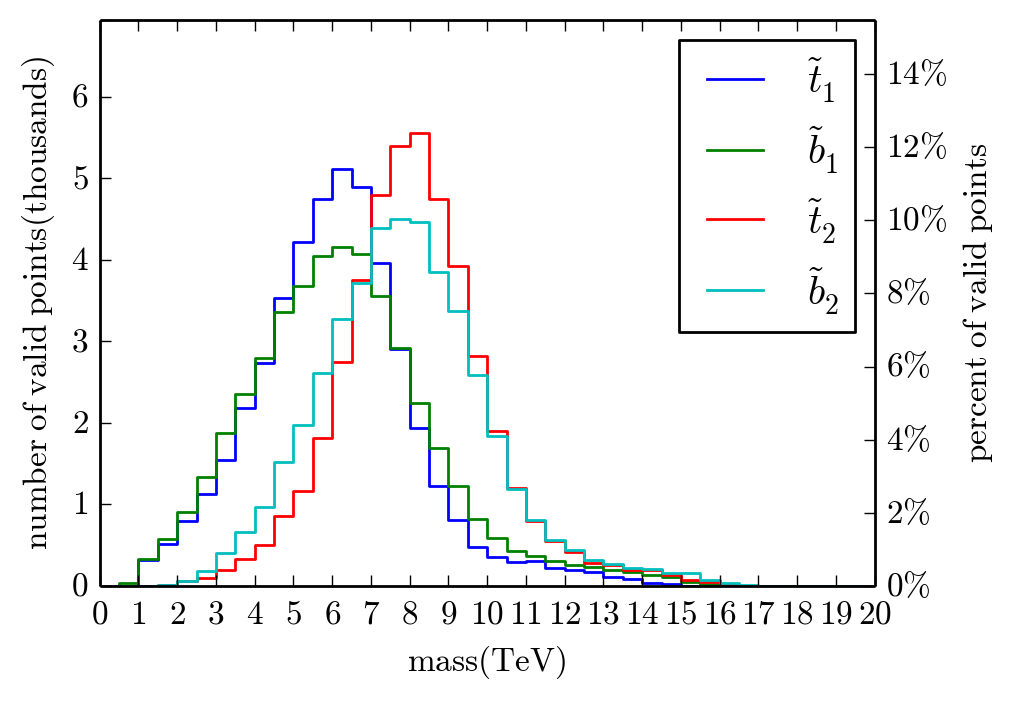}
\caption{\small Histograms of the squark masses from the valid points in the main scan. The first- and second-family left-handed squarks are shown in the top-left panel. Because they come in $SU(2)_{L}$ doublets, and the first- and second-family squarks must be degenerate, all four of these squarks have nearly identical mass and the histograms coincide. The first- and second-family right-handed squarks are shown in the top-right panel. The right-handed down squarks are generally lighter than their up counterparts because of the effect of the $U(1)_{3R}$ charge in the RGEs. The third family squarks are shown in the bottom panel. }
\label{fig:1052}
\end{figure}
\begin{figure}
\centering
	\includegraphics[scale=0.6]{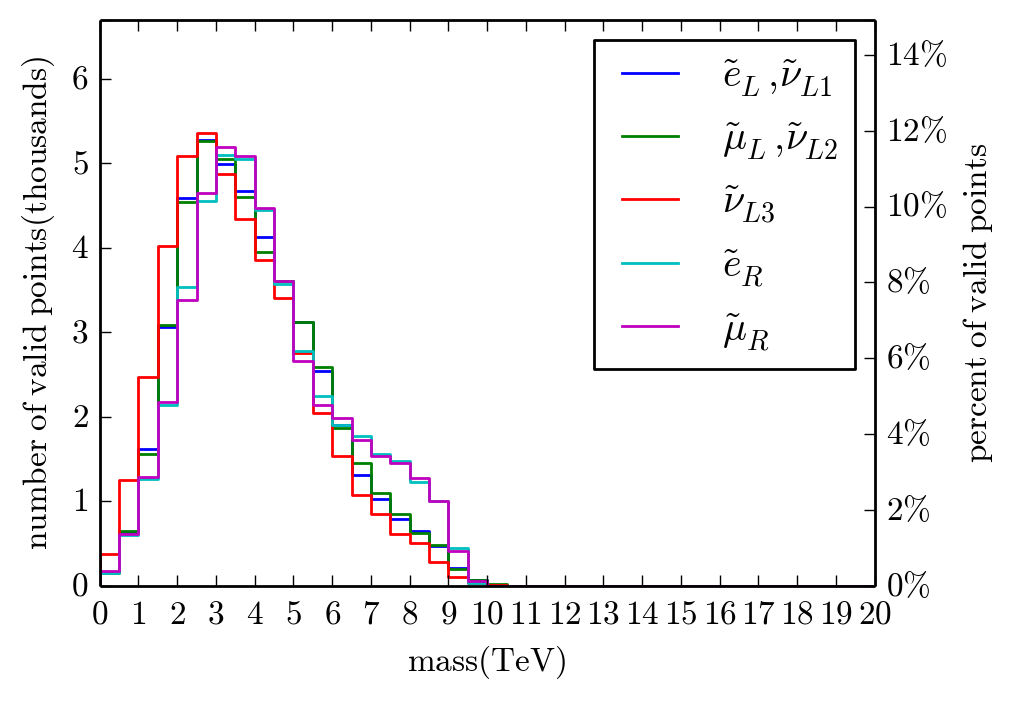}
	\includegraphics[scale=0.6]{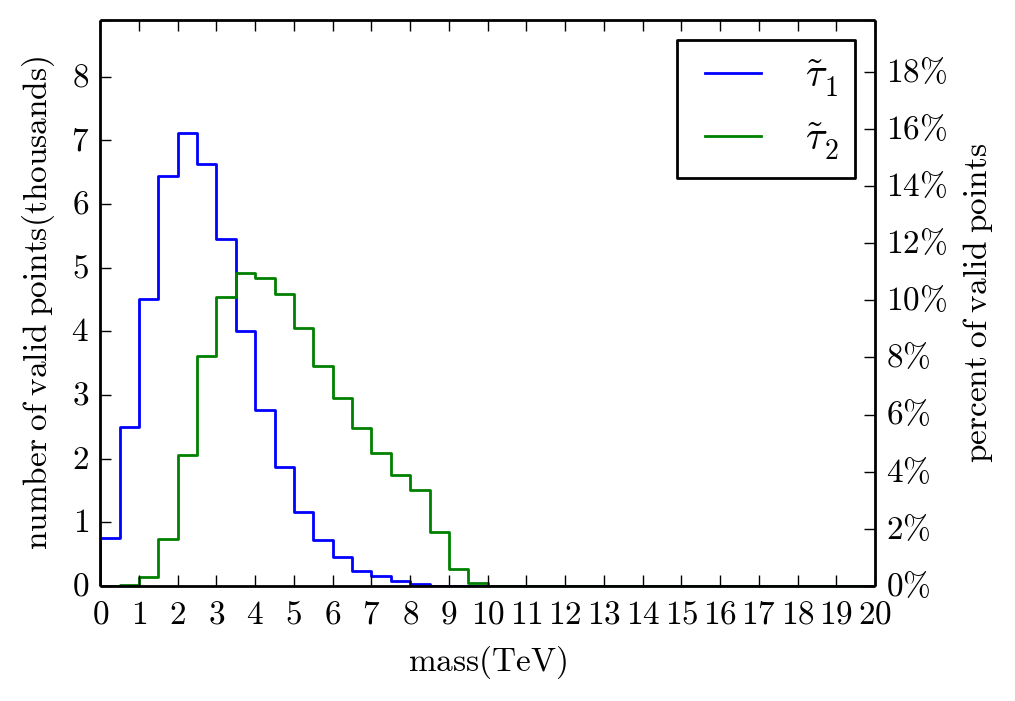}
	\includegraphics[scale=0.6]{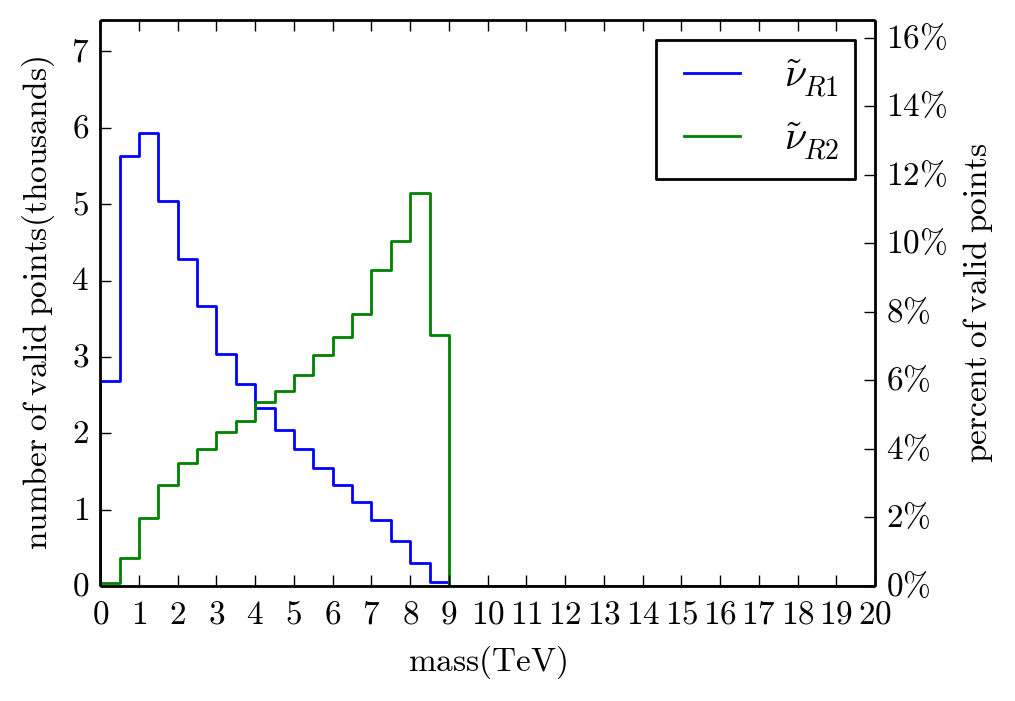}
\caption{\small Histograms of the sneutrino and slepton masses associated with the valid points in the main scan. First- and second-family entries are in the top-left panel, along with the third family left-handed sneutrino. Staus are in the top-right panel with mass-ordered labeling. In the bottom panel, the first- and second-family right-handed sneutrinos are labeled such that $\tilde \nu_{R1}$ is always lighter than $\tilde \nu_{R2}$.}
	\label{fig:hist.sleptons}
\end{figure}
\begin{figure}
\centering
	\includegraphics[scale=0.6]{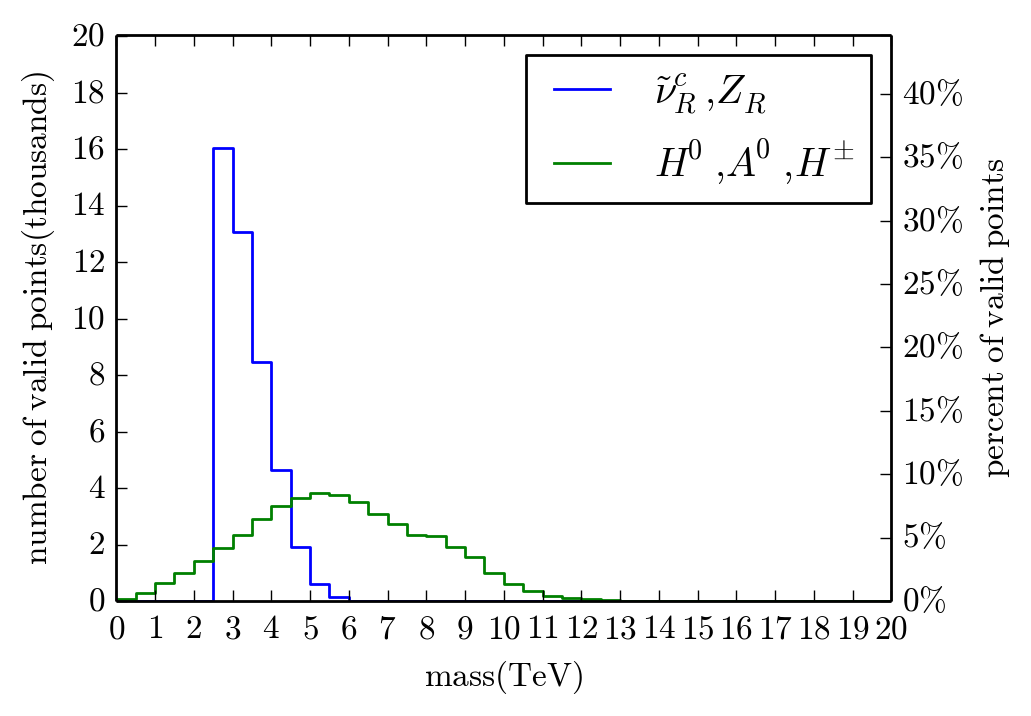}
	\includegraphics[scale=0.6]{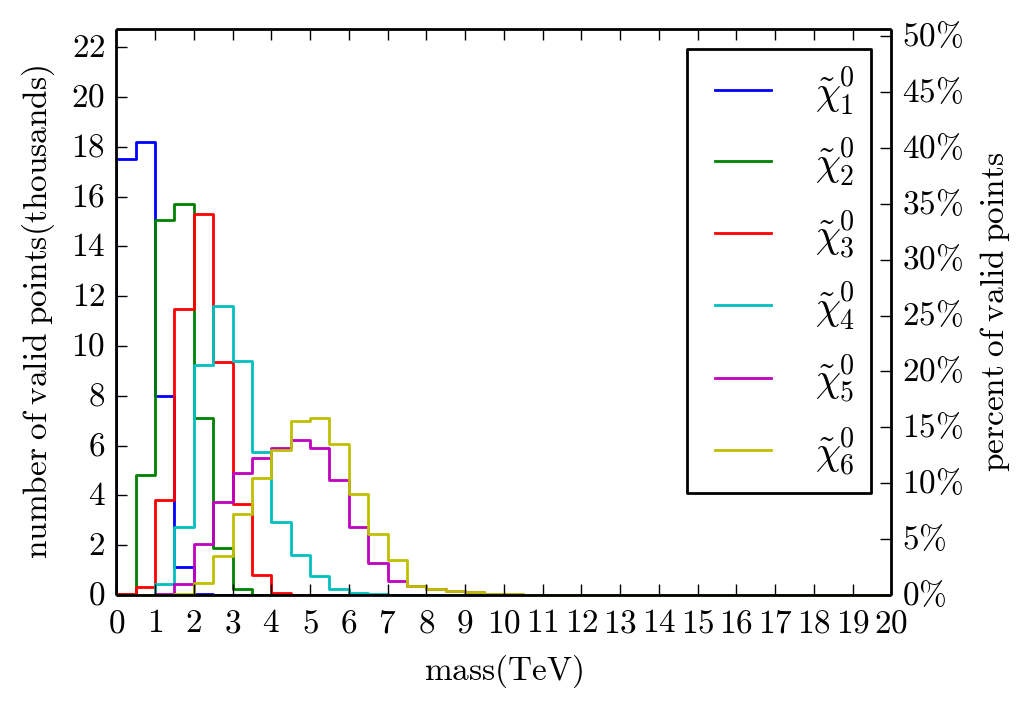}
	\includegraphics[scale=0.6]{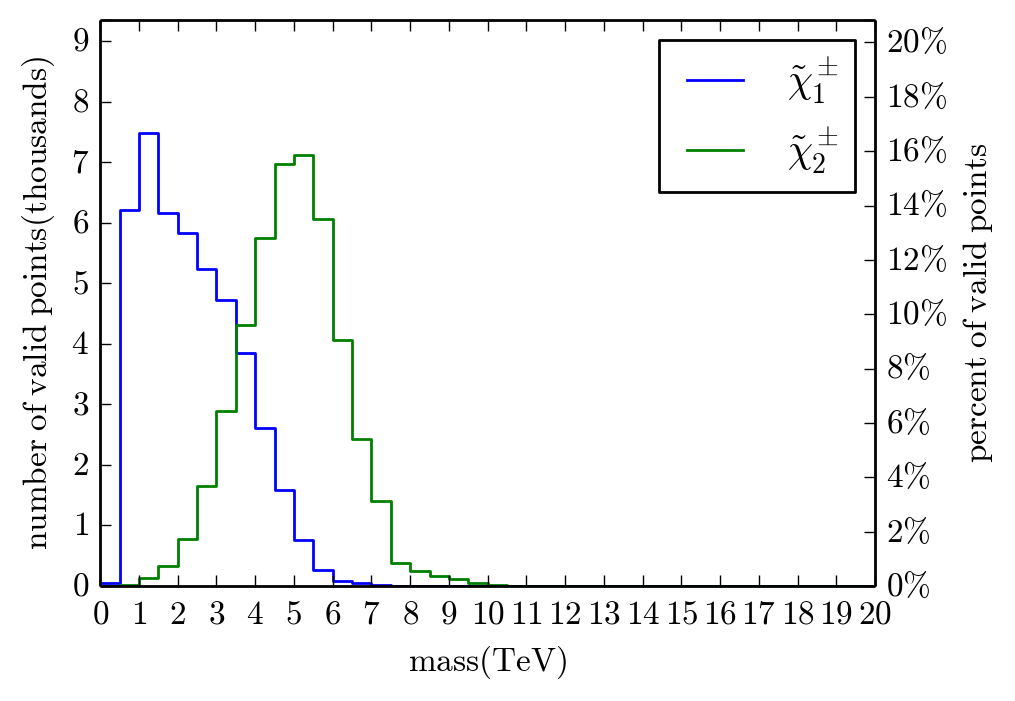}
	\includegraphics[scale=0.6]{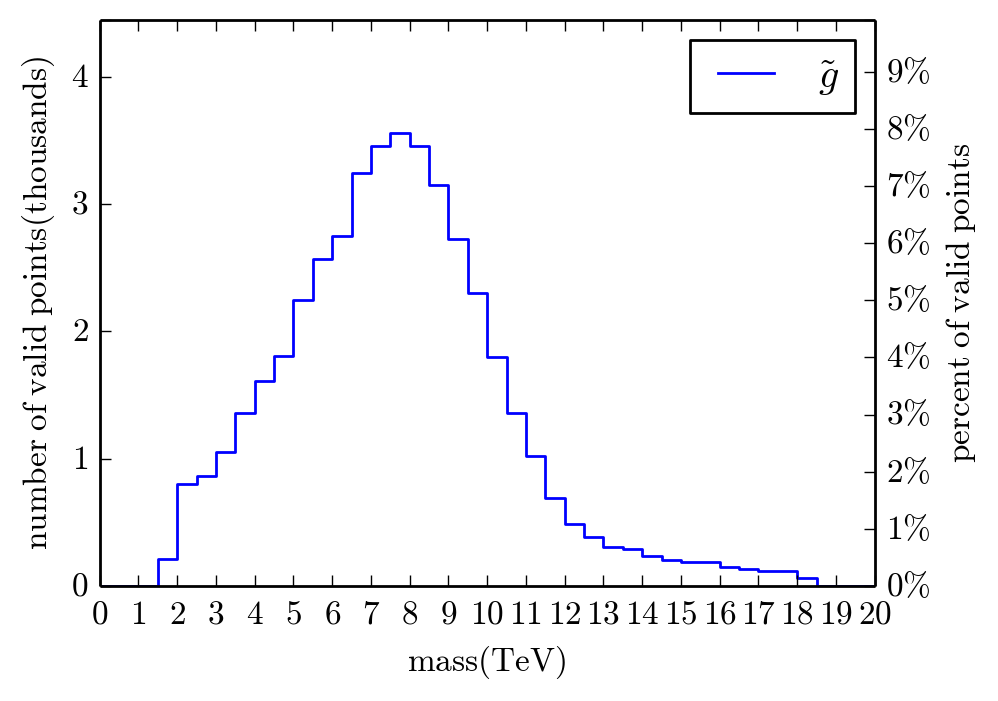}
\caption{\small The CP-even component of the third-family right-handed sneutrino, heavy Higgses, neutralinos, charginos  and the gluino in the valid points from our main scan. The CP-even component of the third-generation right-handed sneutrino is degenerate with $Z_R$. The $\tilde \chi^0_5$ and $\tilde \chi^0_5$ are typically Higgsinos.}
\label{fig:1054}
\end{figure}

Figure~\ref{fig:1052} shows histograms of the squark masses. Because they come in $SU(2)_{L}$ doublets and the first- and second-family squarks must be degenerate, all four of the first- and second-family left-handed squarks have nearly identical mass and the histograms coincide. The degeneracy of first- and second-family squarks is also evident in the right-handed squark masses. The first and second family right-handed down squarks are generally lighter than their up counterparts because of the effect of the $U(1)_{3R}$ charge in the RGEs.
Figure~\ref{fig:hist.sleptons} shows histograms of the masses of the sneutrinos and sleptons. The third-family sleptons and left-handed sneutrinos tend to be the lighter because of the influence of the $\tau$ Yukawa coupling. The right-handed sneutrinos are labeled such that $\tilde \nu_{R_1}$ is always lighter than $\tilde \nu_{R_2}$.
Figure~\ref{fig:1054} presents histograms of the CP-even component of the third-generation right-handed sneutrino, the heavy Higgses, the neutralinos, the charginos, and the gluino. The CP-even component of the third-generation right-handed sneutrino is degenerate with $Z_R$. It is always heavier than 2.5 TeV because we have imposed the collider bound on $Z_R$. The neutralinos and charginos are labeled from lightest to heaviest as is canonical in SUSY models. The $\tilde \chi^0_5$ and $\tilde \chi^0_6$ are typically Higgsinos.
We emphasize that all of the above histograms are calculated using our main scan; that is, for the choice of $M=2700$ GeV and $f=3.3$. We remind the reader that these values were chosen in \cite{{Ovrut:2015uea}} so as to maximize the number of valid points and repeated in this paper so as to enable simple comparison with the split Wilson mass results.  However, the mass scale of these histograms is heavily dependent on the choice of $M$. Smaller (larger) values for $M$ will move the above distributions distinctly toward lighter (heavier) sparticle masses.

\begin{figure}[!htbp]
\centering
	\includegraphics[scale=0.6]{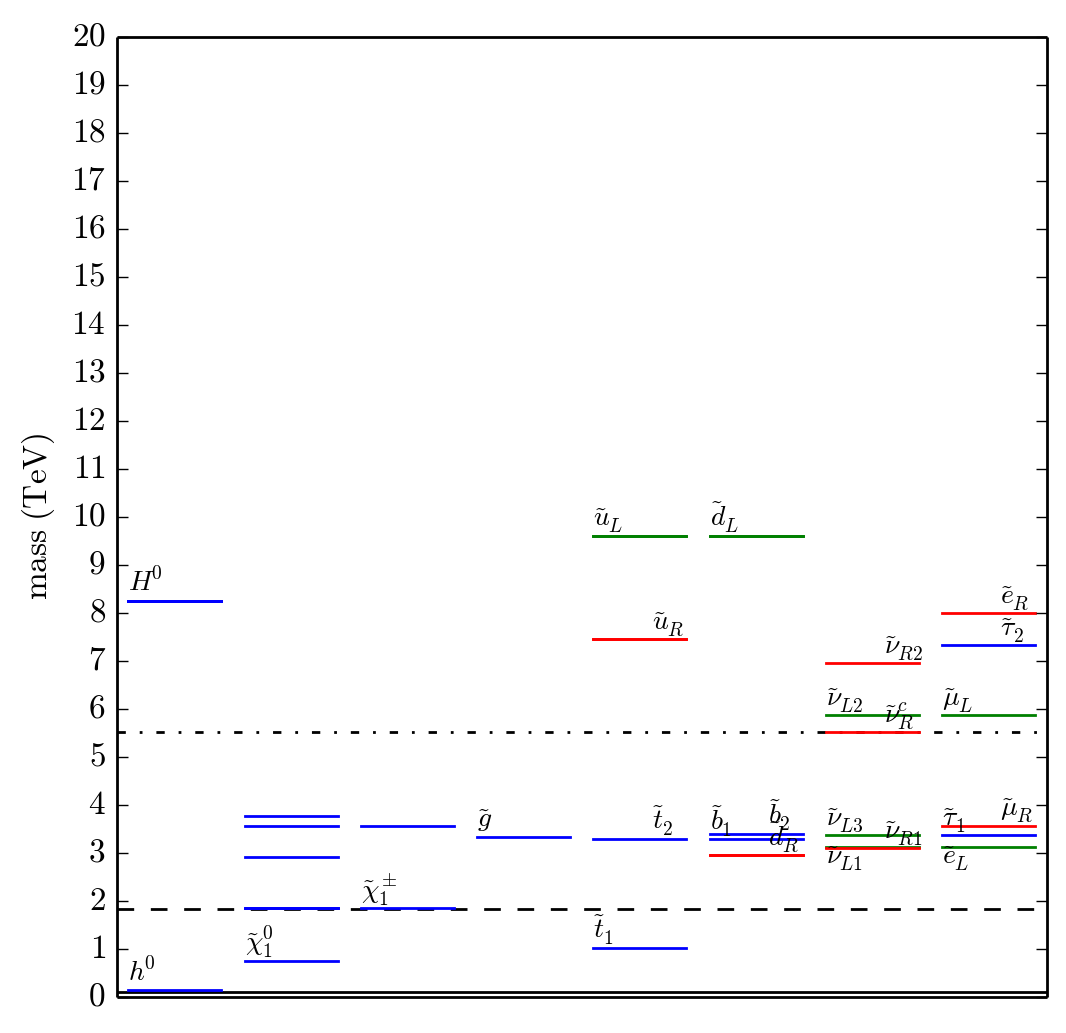}
	\includegraphics[scale=0.6]{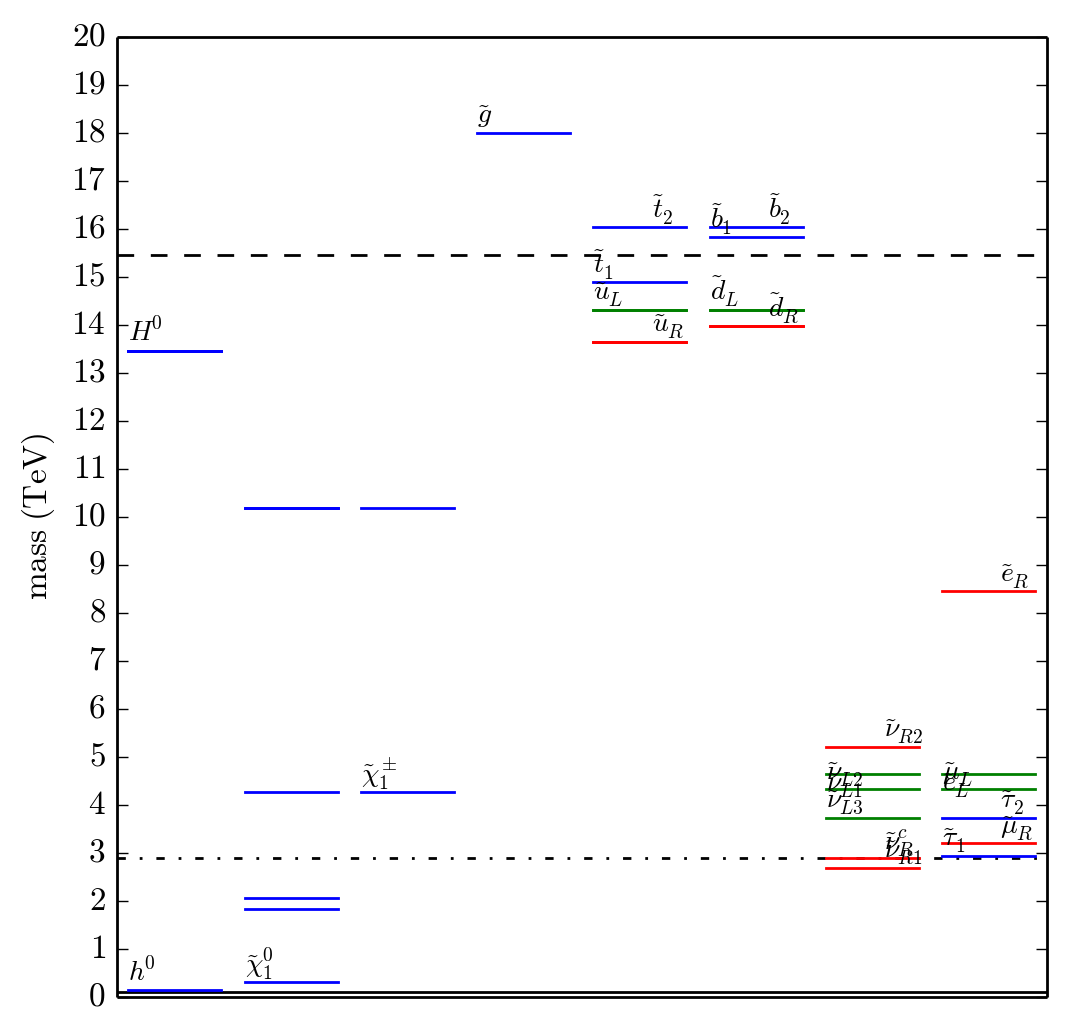}
	\caption{\small Two sample physical spectra with a right-side-up hierarchy and upside-down hierarchy. The $B-L$ scale is represented by a black dot-dash-dot line. The SUSY scale is represented by a black dashed line. The electroweak scale is represented by a solid black line. The label $\tilde u_L$ is actually labeling the nearly degenerate $\tilde u_L$ and $\tilde c_L$ masses. The labels $\tilde u_R$, $\tilde d_L$ and $\tilde d_R$ are similarly labeling the nearly degenerate first- and second- family masses.}
	\label{fig:156}
\end{figure}

\begin{figure}[!htbp]
\centering
	\includegraphics[scale=0.6]{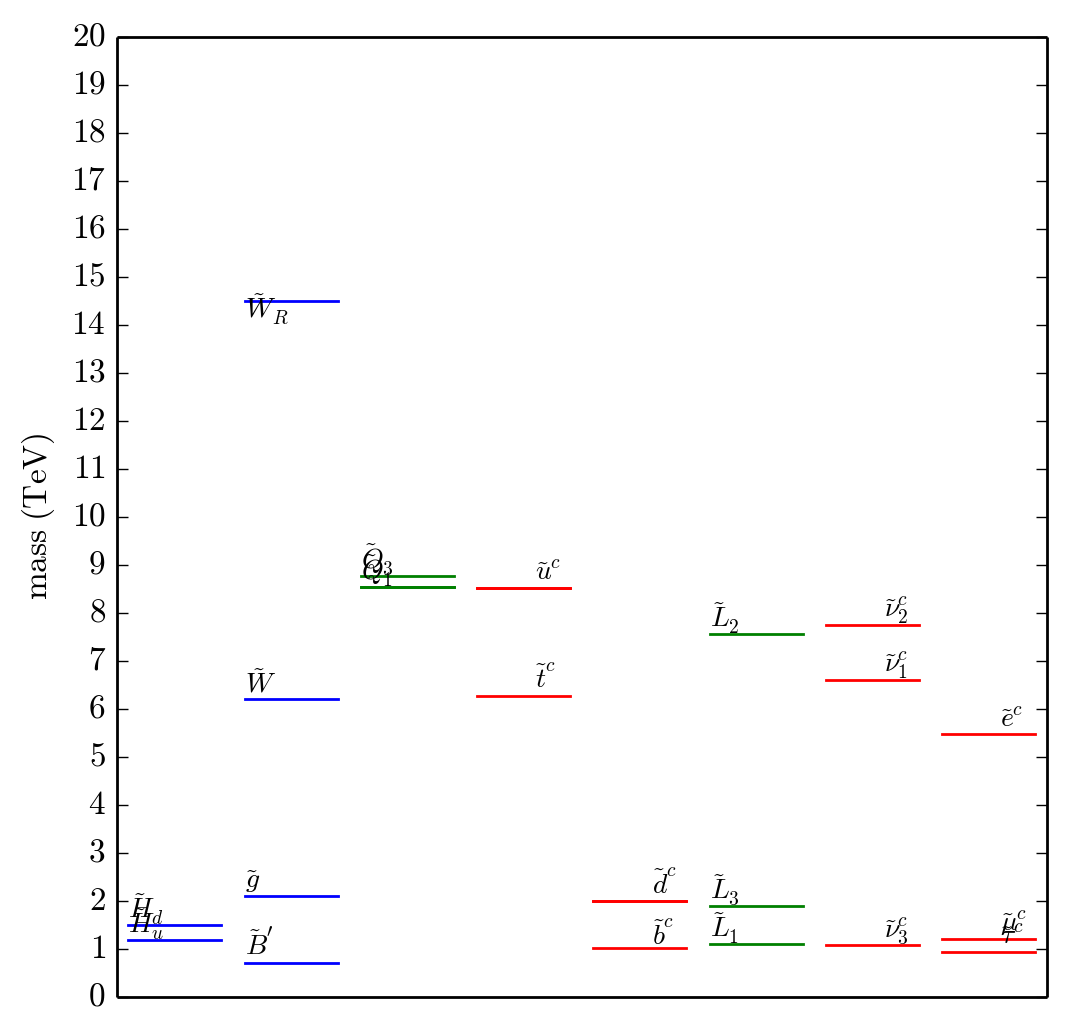}
	\includegraphics[scale=0.6]{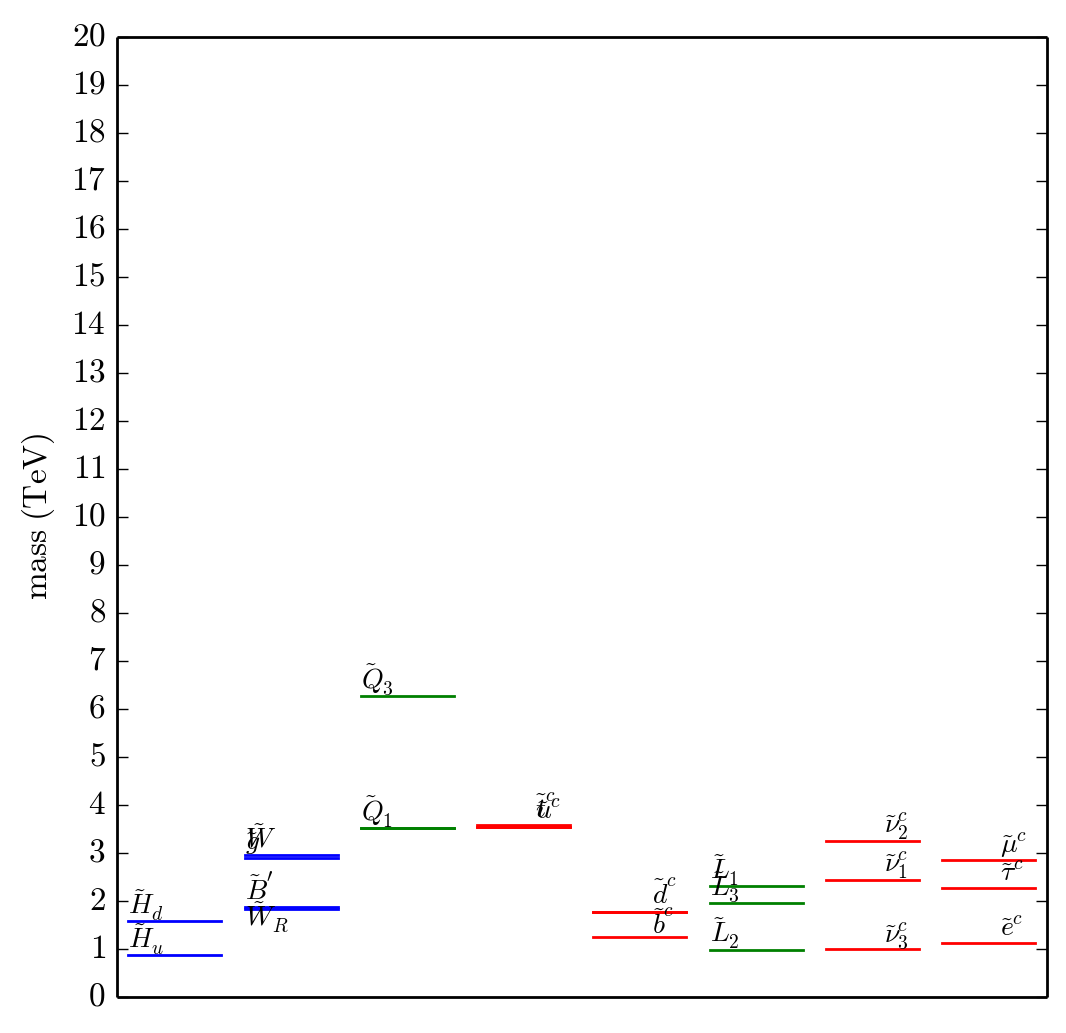}
	\caption{\small Example high-scale boundary conditions at $\left<M_{U}\right>$ for the two valid points with the largest and smallest amount of splitting. The label $\tilde Q_1$ is actually labeling the nearly degenerate $\tilde Q_1$ and $\tilde Q_2$ soft masses. The labels $\tilde u^c$ and $\tilde d^c$ are similarly labeling the nearly degenerate first and second family masses.}
	\label{fig:160}
\end{figure}

Plots of the physical particle spectra for two valid points are presented in Fig.~\ref{fig:156}. 
These two points are selected from the pool of valid points from the main scan based on the simple criteria that they are the valid points with the largest right-side-up and upside-down hierarchy respectively; that is, the largest splittings between the $B-L$ and SUSY scales in the two possible hierarchies.
Plots of the high-scale boundary values for two sample valid points from our main scan are presented in Fig.~\ref{fig:160}. While these look like Figs.~\ref{fig:156}, they do not correspond to physical masses but, rather, mass parameters at $\left<M_{U}\right>$. These two valid points are selected from the pool of valid points from the main scan based on a simple criterion. The two plots show the valid points with the largest and smallest amount of splitting in the initial values of the scalar soft mass parameters. The amount of splitting is defined as the standard deviation of the initial values of the 20 scalar soft mass parameters.

%

\section{Fine-Tuning}


\begin{figure}
\centering
\includegraphics[scale=1.2]{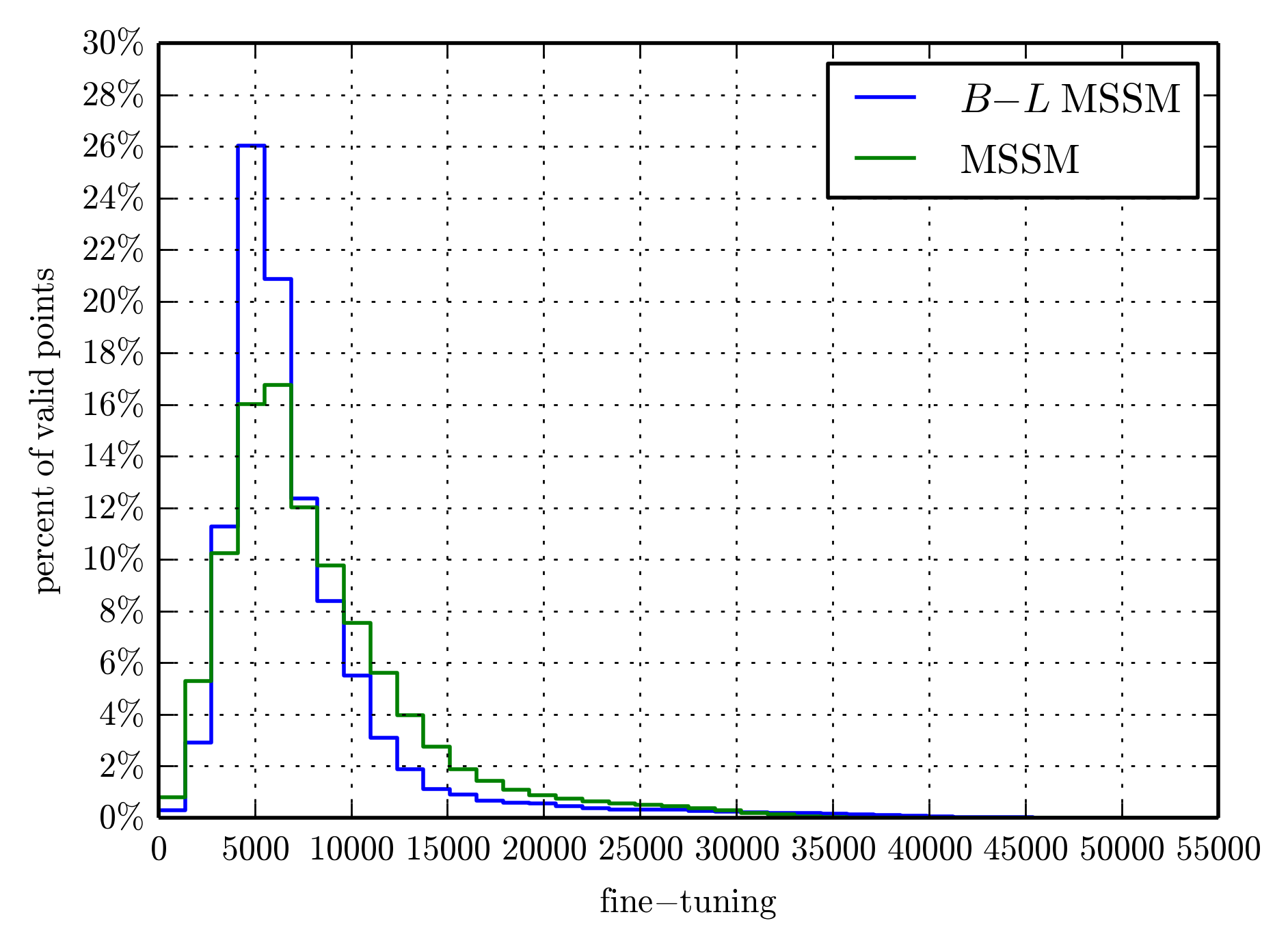}
        \caption{The blue line in the histogram shows the amount of fine-tuning required for valid points in the main scan of the simultaneous Wilson line $B-L$ MSSM. Similarly, the green line specifies the amount of fine-tuning necessary for the valid points of the R-parity conserving MSSM--computed using the same statistical procedure as for the $B-L$ MSSM with $M=2700$ GeV and $f=3.3$. The $B-L$ MSSM shows slightly less fine-tuning, on average, than the MSSM.}
\label{fig:fineTuning}
\end{figure}

A detailed discussion of the the little hierarchy problem, fine-tuning and the Barbieri-Giudice (BG) method of quantifying the degree of fine-tuning was presented in \cite{Ovrut:2015uea}. Here, we simply give the results in the simultaneous Wilson line scenario discussed in this paper. Unlike the quantitites presented above, which can differ substantially from the split Wilson line results in \cite{Ovrut:2015uea}, the BG fine-tuning histogram for simultaneous Wilson lines in the $B-L$ MSSM is very similar to that of the split Wilson line scenario. Be that as it may, for completeness, we present it here--along with the fine-tuning histogram for the R-parity conserving MSSM--in Fig.~\ref{fig:fineTuning}. Note that the highest percentage of valid points require fine-tuning of the order of 1/4,000 -- 1/5,000. However, there remain a small number of points with fine-tuning less than 1/1,000. As in the split Wilson line scenario, the simultaneous Wilson line $B-L$ MSSM manifests somewhat less fine-tuning than the R-parity conserving MSSM.


\section{String Threshold Corrections}


As discussed in the Introduction and Section III, and graphically illustrated for a valid initial point in Figure \ref{fig:c}, the four gauge couplings of the $B-L$ MSSM do not unify at $\left<M_{U}\right>$ for simultaneous Wilson line masses; that is, when 
\begin{equation}
 \left<M_{U}\right> =  M_{\chi_{3R}} = M_{\chi_{B-L}}  \ .
 \label{hani2}
 \end{equation}
However, as described in the Introduction, the $B-L$ MSSM arises on the observable orbifold plane of heterotic $M$-theory compactified on a Schoen Calabi-Yau threefold with $\pi_{1}={\mathbb{Z}}_{3} \times {\mathbb{Z}}_{3}$ and a holomorphic vector bundle with $SU(4) \subset E_{8}$ structure group. That is, the $B-L$ MSSM is a low energy effective theory of heterotic string theory. Hence, as discussed in numerous papers \cite{Kaplunovsky:1992vs, Kaplunovsky:1995jw, Mayr:1993kn, Dienes:1995sq, Dienes:1996du, Dolan:1992nf, Kiritsis:1996dn, Nilles:1997vk, Nilles:1998uy, Ghilencea:2001qq, Klaput:2010dg, deAlwis:2012bm, Bailin:2014nna}, it is expected that at {\it string tree level} all four gauge couplings, along with the dimensionless gravitational parameter 
\begin{equation}
\sqrt{8\pi\frac{G_{N}}{\alpha^{\prime}}} 
\label{hani3}
\end{equation}
where $G_{N}$ is Newton's constant and $\alpha^{\prime}$ is the string Regge slope, unify to a single parameter $g_{\rm string}$ at a ``string unification'' scale
\begin{equation}
M_{\rm string}=g_{\rm string} \times 5.27 \times 10^{17}~ \mbox{GeV} \ .
\end{equation}
The string coupling parameter $g_{\rm string}$ is set by the value of the dilaton, and is typically of ${\cal{O}}(1)$. A common value in the literature, see for example \cite{Dienes:1996du,Bailin:2014nna,Nilles:1998uy}, is $g_{\rm string}= 0.7$ which, for specificity, we will use henceforth. Therefore, we take $\alpha_{\rm string}$ and the string unification scale to be 
\begin{equation}
\alpha_{\rm string}=\frac{g_{\rm string}^{2}}{4\pi} = 0.0389 , \quad    M_{\rm string}=3.69 \times 10^{17}~ \mbox{GeV} 
\label{hani4}
\end{equation}
respectively. Note that $ M_{\rm string}$ is approximately an order of magnitude larger than $\left<M_{U}\right>$. Below $M_{\rm string}$ however, the couplings begin to evolve according to the RGEs of effective field theory. This adds another--fourth-- scaling regime to the three discussed at the beginning of Section IV. This new regime is

\begin{figure}
\centering
\includegraphics[scale=.75]{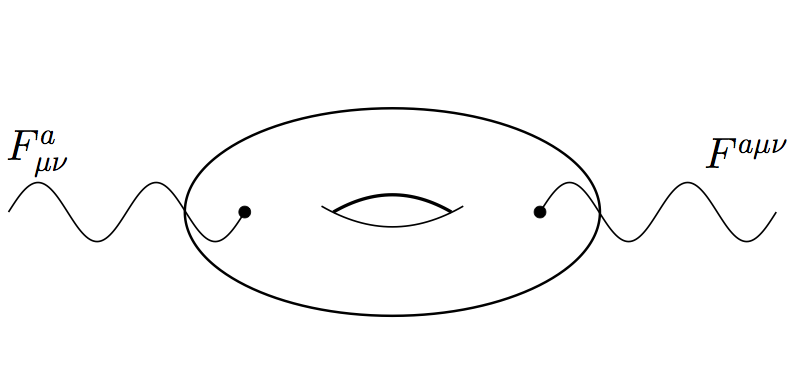}
\caption{The worldsheet correlation function $\left<F^{a}_{\mu\nu}F^{a\mu\nu}\right>$ on the genus-one string worldsheet is a typical example of heavy string threshold correction terms that need to be calculated.}
\label{fig:rehan}
\end{figure}

\begin{itemize} 

\item $M_{\rm string}$ -- $\left<M_{U}\right>$: The effective field theory in this regime remains that of the $B-L$ MSSM with the couplings ${\alpha}_{a}$,  $a=3,2,3R, BL^{\prime}$ and the slope factors  
\begin{equation}
		b_3 = -3~,~b_2 = 1~, ~b_{3R}= 7~,~ b_{BL'}= 6  
\label{hani5}
\end{equation}
as in Eqn.~\eqref{red1}.
However, the RGEs are now altered to become \footnote{The RGE for the a-th gauge coupling generically contains the term $k_{a} \alpha_{\rm string }^{-1}$ on the right-hand side, where $k_{a}$, $a=3,2,3R,BL^{\prime}$ are the associated string affine levels. However, these are all unity for the scaled gauge couplings of the $B-L$ MSSM.}
\begin{equation}
4\pi {\alpha_{a}}^{-1}( p)=4\pi \alpha_{\rm string }^{-1}-b_{a}\ln\big(\frac{p^2}{M_{\rm string}^{2}}\big) + {\tilde{\Delta}}_{a}.
\label{hani6}
\end{equation}
Note that the one-loop running couplings no longer unify exactly at $M_{\rm string}$. Rather, they are ``split'' by dimensionless threshold effects. These arise predominantly from massive genus-one string modes that contribute to the correlation function $\left<F^{a}_{\mu\nu}F^{a  \mu\nu}\right>$ and, hence, to the ${\alpha}_{a}$ gauge couplings. This is depicted graphically in Figure \ref{fig:rehan}.

\end{itemize}

\begin{figure}
\centering
	\includegraphics[scale=1.2]{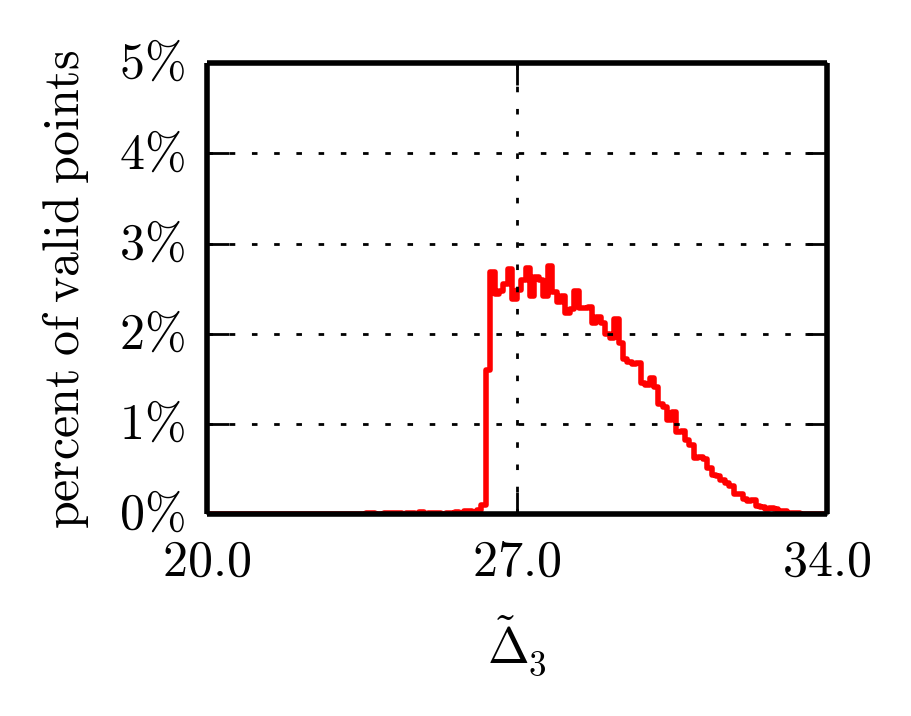}
	\includegraphics[scale=1.2]{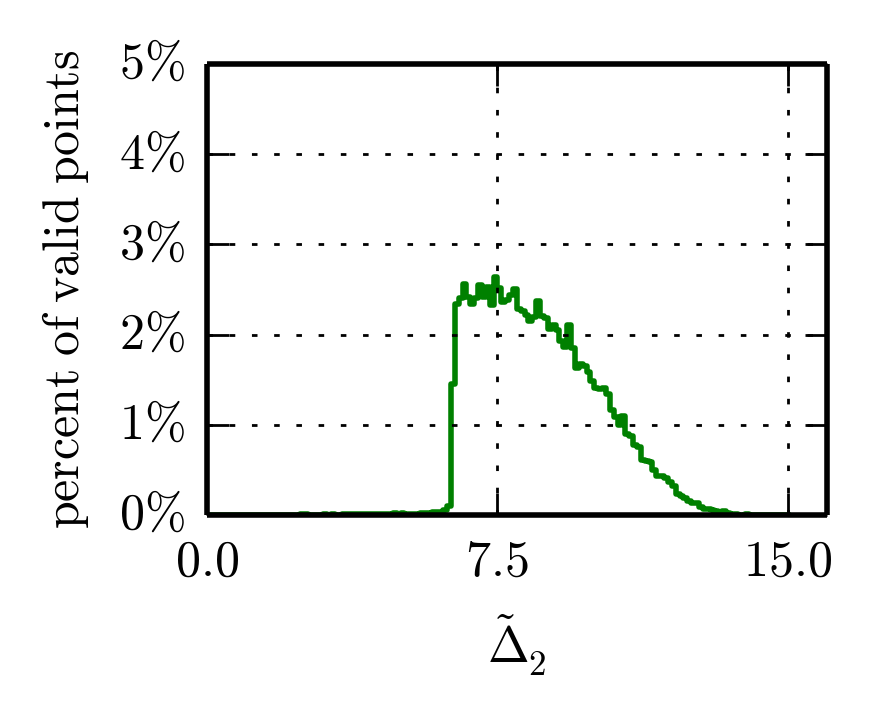}
	\includegraphics[scale=1.2]{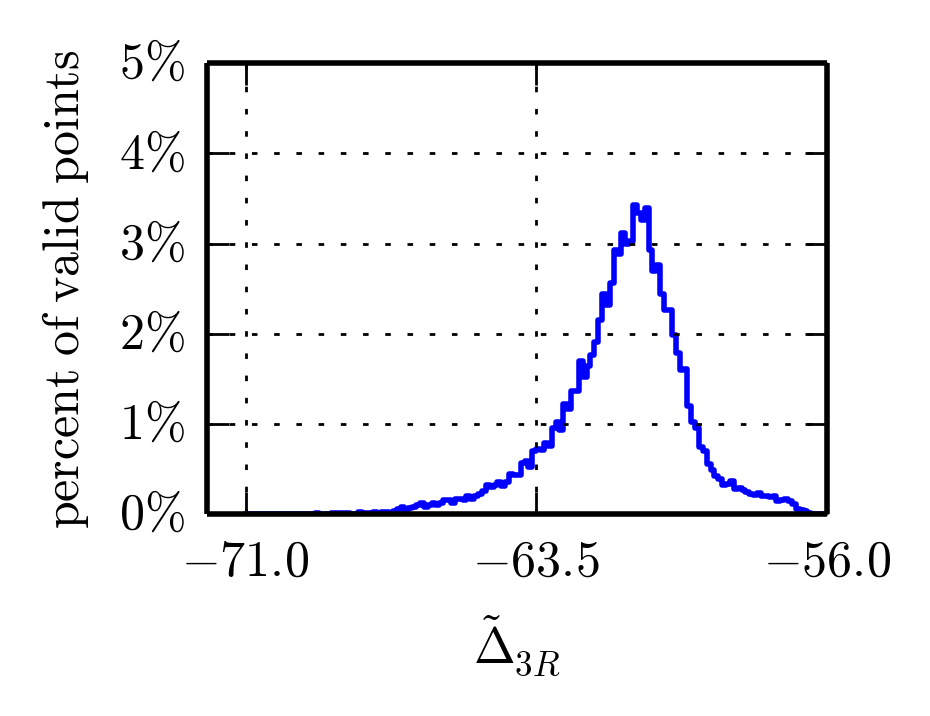}
	\includegraphics[scale=1.2]{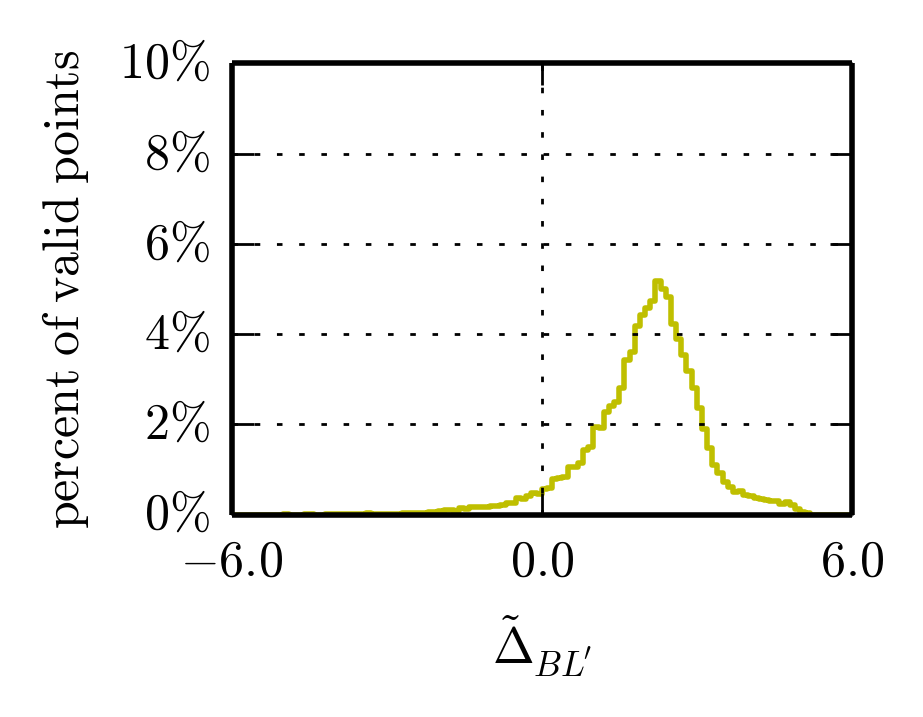}
\caption{\small Histograms of each of the heavy string thresholds ${\tilde{\Delta}}_{a}$, $a=3,2,3R,BL'$ arising from the $44,884$ phenomenologically valid points of our statistical survey. Each threshold value is plotted against the percentage of valid points giving rise to it. The bin width is 0.1.}
	\label{fig:thresh1}
\end{figure}

\begin{figure}
\centering
	\includegraphics[scale=1.2]{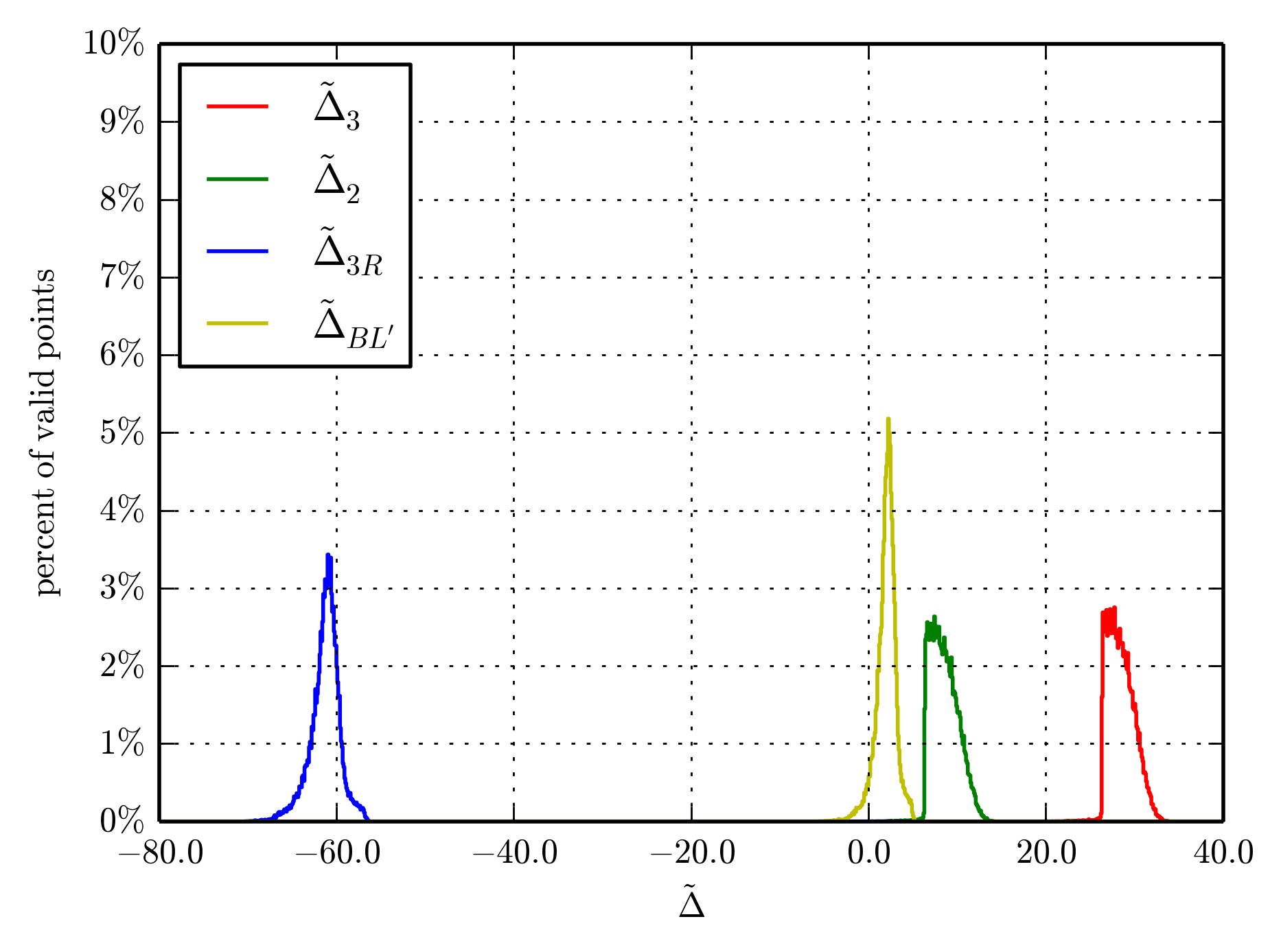}
	\caption{\small All four histograms in Fig. \ref{fig:thresh1} combined into a single graph to elucidate their relative occurrence and values.}
	\label{fig:thresh2}
\end{figure}

Recall that in this paper we have found $44,884$ valid initial points in the space of soft supersymmetry breaking dimensionful couplings-- each of which satisfies all low energy phenomenological criteria. For {\it each} of these points, we can calculate--by scaling from the electroweak scale to $\left<M_{U}\right>$-- the four gauge couplings  
$\alpha_{3}(\left<M_{U}\right>),~ \alpha_{2}(\left<M_{U}\right>), ~\alpha_{3R}(\left<M_{U}\right>)$ and $ \alpha_{BL}^{\prime}(\left<M_{U}\right>)$. Note that in this analysis, we have defined an ``average'' SUSY scale $M_{SUSY}$, the $B-L$ breaking scale $M_{B-L}$, as well as an ``average'' unification scale $\left<M_{U}\right>$, in (28), (24) and (22) respectively. The RGEs have been scaled through the requisite intermediate regimes with the appropriate beta-function coefficients. That is, we have already taken into account the predominant threshold effects associated with each of these scales. For statistically ``average'' valid initial points--the vast majority of the phenomenologically acceptable initial soft SUSY breaking parameters--possible additional threshold effects arising from the ``splitting'' of particle masses around these scales are expected to be relatively small--and will be systematically ignored relative to the heavy string thresholds. That is, the ${\tilde{\Delta}}_{a}$ , $a=3, 2, 3R, BL^{\prime}$
parameters in (50) will closely approximate the four heavy string gauge thresholds. With this input, using eqn.~\eqref{hani5}, $p=\left<M_{U}\right>=3.15 \times 10^{16}~$GeV and $\alpha_{\rm string}$,  $M_{\rm string}$ given in \eqref{hani4}, one can calculate the associated heavy string thresholds from~\eqref{hani6}; that is,
\begin{equation}
 {\tilde{\Delta}}_{a}=4\pi  {\alpha_{a}}^{-1}(\left<M_{U}\right> )-4\pi \alpha_{\rm string }^{-1}+b_{a}\ln\big(\frac{\left<M_{U}\right>^2}{M_{\rm string}^{2}}\big) \ .
\label{hani7}
\end{equation}
for each $a=3,2,3R, BL^{\prime}$. Of course, these thresholds are expected to differ for each different valid initial point. It follows that one should analyze the thresholds statistically--graphing the dispersion of each as one runs over the $44,884$ valid initial points. The histograms associated with each of these four thresholds 
are presented in Figure~\ref{fig:thresh1}.
To better understand the relationship of these different thresholds, we find it useful to plot all four of them in a single histogram. This is presented in Figure~\ref{fig:thresh2}.

\begin{figure}
\centering
	\includegraphics[scale=1.0]{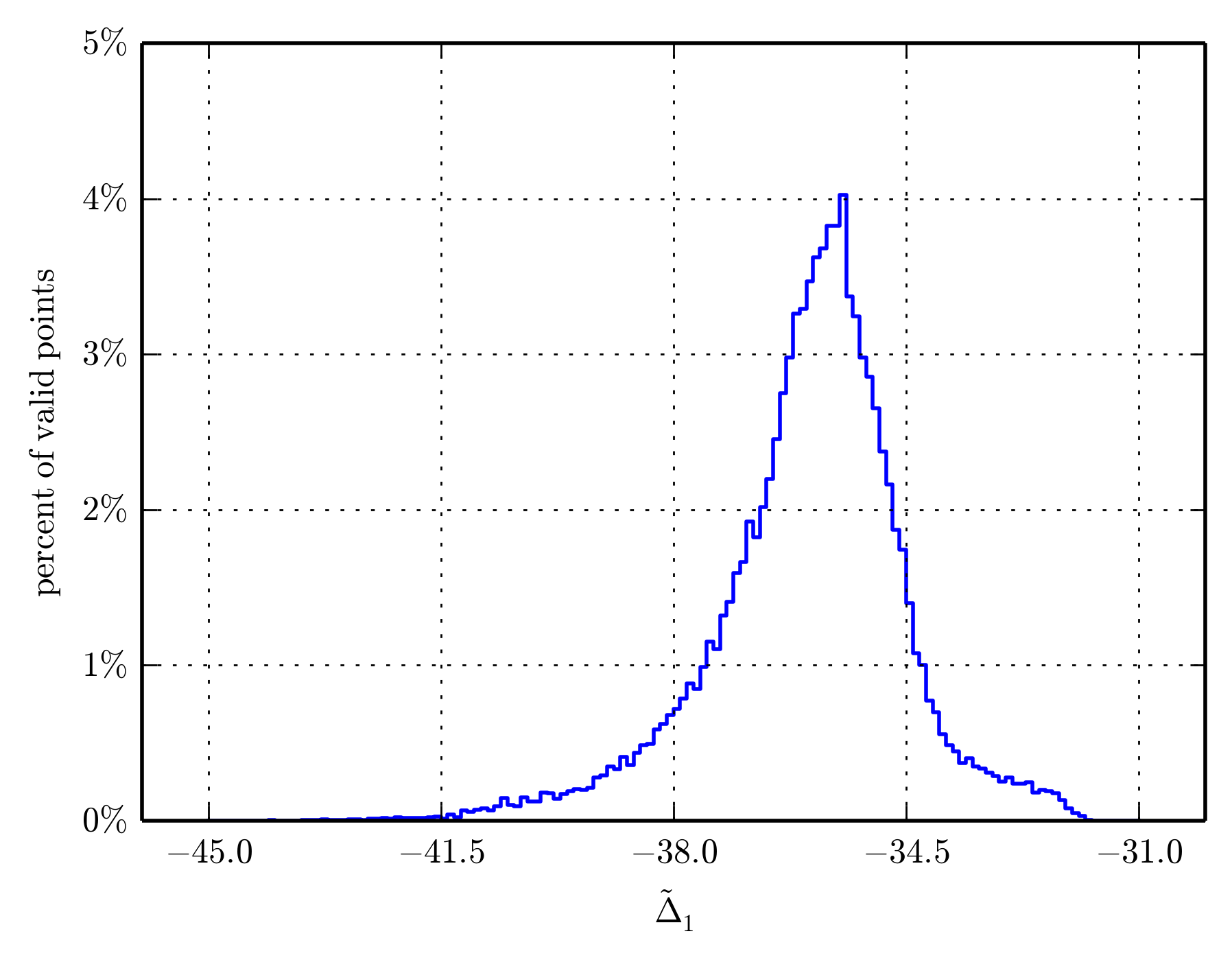}
	\caption{\small Histogram of the string hypercharge threshold ${\tilde{\Delta}}_{1}$ arising from the $44,884$ phenomenologically valid points of our statistical survey. Each threshold value is plotted against the percentage of valid points giving rise to it. The bin width is 0.1.}
	\label{fig:thresh3}
\end{figure}

\begin{figure}
\centering
	\includegraphics[scale=1.2]{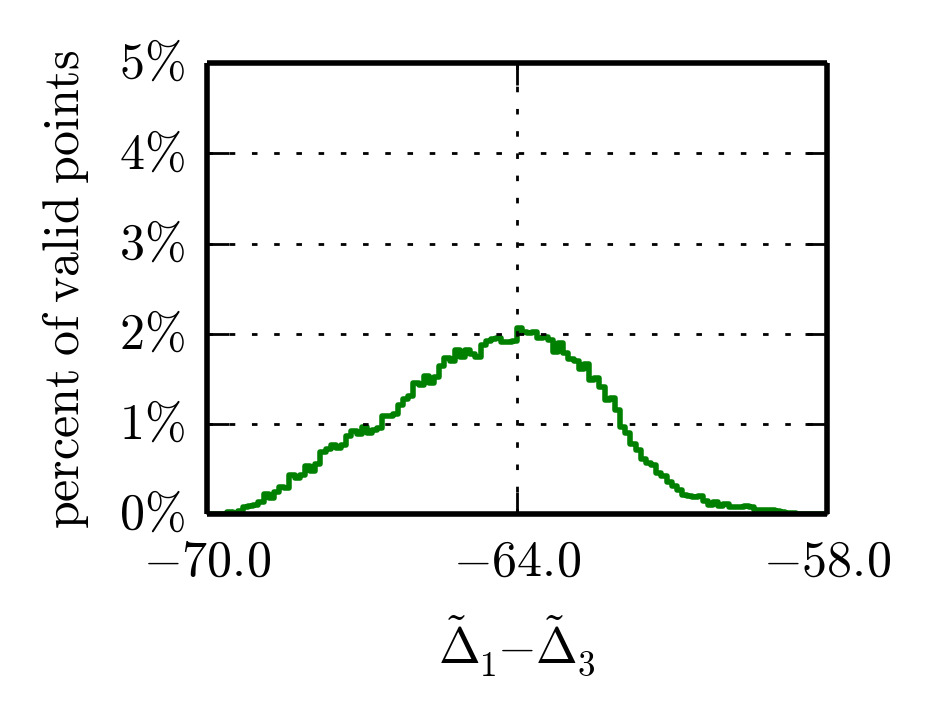}
	\includegraphics[scale=1.2]{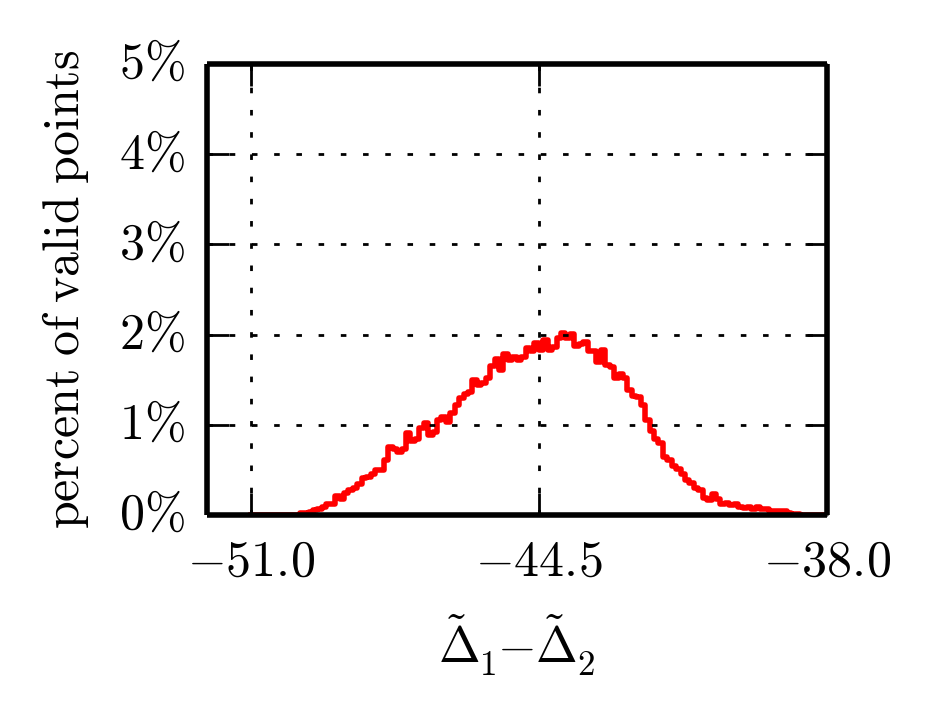}
	\includegraphics[scale=1.2]{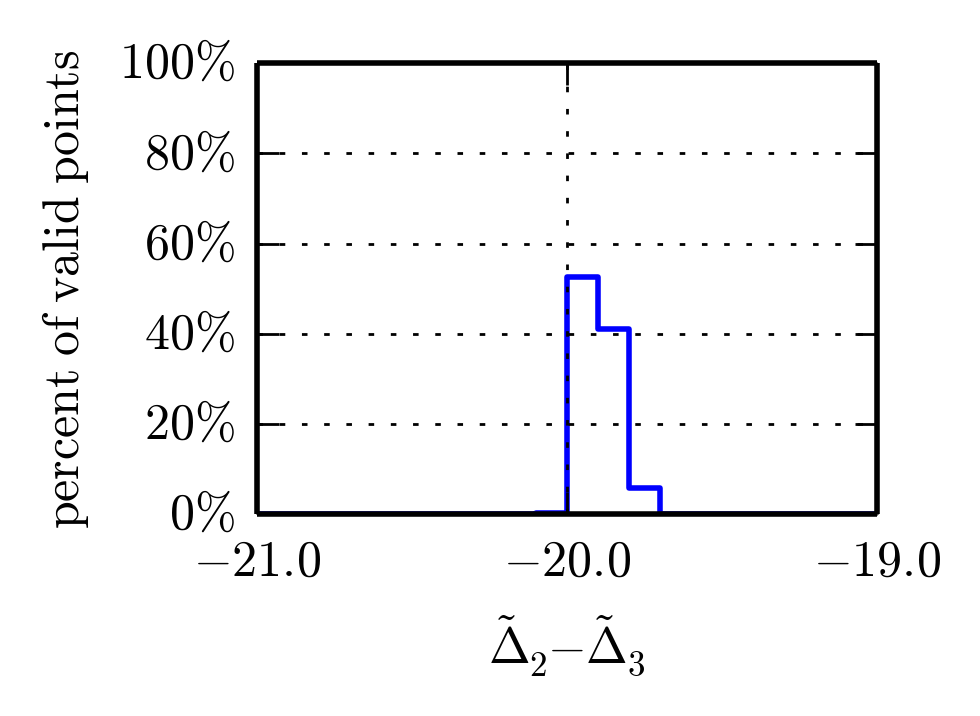}
\caption{\small Histograms of our statistical predictions for the values of ${\tilde{\Delta}}_{1}-{\tilde{\Delta}}_{2}$, ${\tilde{\Delta}}_{1}-{\tilde{\Delta}}_{3}$, and ${\tilde{\Delta}}_{2}-{\tilde{\Delta}}_{3}$. The third of these plots looks different because the quantity $\tilde\Delta_2-\tilde\Delta_3$ falls in a very narrow range. The bin width in all three plots is 0.1.}
	\label{fig:thresh4}
\end{figure}

 It is also useful to calculate the string threshold associated with the Abelian hypercharge coupling $\alpha_{1}$ defined, using \eqref{eq:1.3R.BL} and \eqref{home1}, by \footnote{As with the other $B-L$ MSSM gauge couplings, this scaled hypercharge coupling has string affine level $k_{1}=1$.}
 \begin{equation}
 \alpha^{-1}_{1}=\frac{3}{5}\alpha^{-1}_{3R}+\frac{2}{5} \alpha^{-1}_{BL'} \ .
 \label{rehan2}
 \end{equation}
 The associated statistical histogram is given in Figure \ref{fig:thresh3}.
It is well-known \cite{Kaplunovsky:1992vs,Dienes:1996du,Dolan:1992nf,Mayr:1993kn} that each string threshold breaks into two parts, 
\begin{equation}
{\tilde{\Delta}}_{a}= {\mathbb{Y}}+\Delta_{a} \ ,
\label{rehan3}
\end{equation}
where $ {\mathbb{Y}}$ is a ``universal'' piece independent of the gauge group and $\Delta_{a}$ records the contributions of all massive string states as they propagate around the genus-one string worldsheet torus diagram shown in Figure \ref{fig:rehan}. Note again that all string affine levels are unity in our normalization. As discussed in \cite{Dienes:1996du}, an explicit calculation of the universal piece $\mathbb{Y}$ is difficult due to the presence of infrared divergences. However, the $\Delta_{a}$ threshold terms, although moduli dependent, can be directly calculated from string theory using a formulation given by V. Kaplunovsky in \cite{Kaplunovsky:1992vs} and by  Kaplunovsky and Louis in \cite{Kaplunovsky:1995jw}. Such calculations are heavily model dependent \cite{Dienes:1995sq,Klaput:2010dg,Bailin:2014nna,Kiritsis:1996dn} and, to date, have not been carried out in the $B-L$ MSSM context. Be that as it may, it is useful to present our experimental predictions for ${\tilde{\Delta}}_{1}-{\tilde{\Delta}}_{2}$, ${\tilde{\Delta}}_{1}-{\tilde{\Delta}}_{3}$, and ${\tilde{\Delta}}_{2}-{\tilde{\Delta}}_{3}$ --from which information about $\Delta_{3}$, $\Delta_{2}$ and $\Delta_{1}$ can be inferred. The statistical results for these three quantities are presented in Figure \ref{fig:thresh4}.
It would be very interesting to compare these results to direct calculations using \cite{Kaplunovsky:1992vs,Kaplunovsky:1995jw} in the $B-L$ MSSM context. We will not attempt that in the present paper.

\section{Conclusion}

Our previous work on the $B-L$ MSSM used the constraint of exact gauge unification. That is, we chose $M_U\simeq M_{\chi_{B-L}} > M_{\chi_R}$ and, by appropriately separating the Wilson line masses, forced the gauge couplings to unify exactly at the scale $M_U$. This hypothesis enhanced the specificity of the calculation. For example, it set all gauge couplings to a unified value $\alpha_{u}$ at $M_{U}$  and determined $\sin^2\theta_R$ at the SUSY transition mass. However, the splitting of the Wilson lines limits the analysis to a restricted region of CY moduli space where the associated two-cycles have considerably different radii. Furthermore, it introduces a new scaling regime between $M_{U} \simeq  M_{\chi_{B-L}}$ and $M_{\chi_R}$--in our example a ``left-right'' model with a specific spectrum. For soft SUSY breaking masses in the TeV range, this constraint was reasonable since the separation between the Wilson line masses--and, hence, the difference in the two-cycle radii--was less than an order of magnitude. However, if one tries to take larger values for the soft SUSY breaking masses, the difference in the Wilson line masses grows rapidly. For example, for $10^{4}$ TeV soft masses, the Wilson lines must be separated by a factor of $10^{3}$--reducing the calculation to an extremely unnatural region of CY moduli space. For even larger values of the soft masses, the calculation breaks down completely. It follows that if one wishes to discuss the $B-L$ MSSM for large soft SUSY breaking masses--see our discussion below--then it becomes necessary to analyze the theory for the more natural case of simultaneous, or nearly simultaneous, Wilson lines.
Hence, in this paper, we have carried out the analysis of the $B-L$ MSSM under the more natural hypothesis of equal Wilson line scales; that is, we assumed that $M_U\simeq M_{\chi_{B-L}}\simeq M_{\chi_R}$. Although this hypothesis does not allow exact gauge unification, it is well-known--see the previous section--that string threshold corrections can be responsible for  such non-unification, providing a consistent theoretical framework for this analysis.

Our results indicate that a substantial region of the initial soft SUSY breaking parameter space is consistent with all low-energy experimental data--see Figs.~\ref{fig:1204}-\ref{fig:1148}. We also presented our results for important physical quantities; histograms of the LSP species--see Fig.~\ref{fig:1039}--histograms of the sparticle masses--Figs.~\ref{fig:1052}-\ref{fig:1054}--plots of sample mass spectra at the SUSY scale--see Fig.~\ref{fig:156}--and plots of sample initial mass data at the average unification scale--Fig.~\ref{fig:160}. Although different in some details from the previous exact gauge unification analysis, the results presented in this paper share many common features. In addition, they show that all low-energy phenomenological constraints can be satisfied for a wide range of initial soft SUSY breaking parameters over a more natural and generic region of CY moduli space.
The fundamental difference between the calculations in this paper from those of our previous work is that the gauge couplings of the $B-L$ MSSM can no longer unify at an average mass associated with the inverse CY radius. Rather, the gauge couplings remain split at that scale--a splitting that allows one to compute the heavy string threshold corrections due to heavy states on the genus-one string worldsheet. These string threshold corrections and their differences are analyzed statistically, and predict what one should obtain using a a more direct worldsheet formalism. It would be interesting to see the results of such formal string calculations done in the heterotic $B-L$ MSSM context.

Finally, having achieved these more generic results, we are now in a position to arbitrarily raise the mass scale of the soft SUSY breaking parameters--without encountering the above mentioned difficulties. For example, it is now possible to raise the soft SUSY breaking scale up to $10^{12}$ or $10^{13}$ GeV--values consistent with the mass scale in ``split supersymmetry'' theories \cite{ArkaniHamed:2004fb, Giudice:2004tc, ArkaniHamed:2004yi}. As we will show in a subsequent publication, this freedom will allows us to discuss a new theory of inflation within the context of the $B-L$ MSSM. This work will appear elsewhere \cite{inPreparation}.

\section*{Acknowledgments}

We would like to thank Sogee Spinner for helpful conversations. R. Deen and B.A. Ovrut are supported in part by the DOE under contract No. DE-SC0007901. A. Purves was supported by DOE contract No. DE-SC0007901 when much of this work was carried out.


\begin{thebibliography}{99}

\bibitem{Lukas:1997fg} 
  A.~Lukas, B.~A.~Ovrut and D.~Waldram,
  ``On the four-dimensional effective action of strongly coupled heterotic string theory,''
  Nucl.\ Phys.\ B {\bf 532}, 43 (1998)
  doi:10.1016/S0550-3213(98)00463-5
  [hep-th/9710208].
  
  
\bibitem{Lukas:1998yy} 
  A.~Lukas, B.~A.~Ovrut, K.~S.~Stelle and D.~Waldram,
  ``The Universe as a domain wall,''
  Phys.\ Rev.\ D {\bf 59}, 086001 (1999)
  doi:10.1103/PhysRevD.59.086001
  [hep-th/9803235].
  
\bibitem{Lukas:1998tt} 
  A.~Lukas, B.~A.~Ovrut, K.~S.~Stelle and D.~Waldram,
  ``Heterotic M theory in five-dimensions,''
  Nucl.\ Phys.\ B {\bf 552}, 246 (1999)
  doi:10.1016/S0550-3213(99)00196-0
  [hep-th/9806051].
  
\bibitem{Donagi:1999ez} 
  R.~Donagi, B.~A.~Ovrut, T.~Pantev and D.~Waldram,        
  ``Standard models from heterotic M theory,''
  Adv.\ Theor.\ Math.\ Phys.\  {\bf 5}, 93 (2002)
  [hep-th/9912208].
  

  
\bibitem{Ovrut:1979pk} 
  B.~A.~Ovrut and H.~J.~Schnitzer,
  ``A New Approach To Effective Field Theories,''    
  Phys.\ Rev.\ D {\bf 21}, 3369 (1980).
  doi:10.1103/PhysRevD.21.3369
 
    

\bibitem{Font:1989ai} 
  A.~Font, L.~E.~Ibanez and F.~Quevedo,
  ``Does Proton Stability Imply the Existence of an Extra Z0?,''
  Phys.\ Lett.\ B {\bf 228}, 79 (1989).

\bibitem{Martin:1997ns} 
  S.~P.~Martin,
  ``A Supersymmetry primer,''
  Adv.\ Ser.\ Direct.\ High Energy Phys.\  {\bf 21}, 1 (2010)
  [hep-ph/9709356].

\bibitem{Martin:1992mq} 
  S.~P.~Martin,
  ``Some simple criteria for gauged R-parity,''
  Phys.\ Rev.\ D {\bf 46}, 2769 (1992)
  [hep-ph/9207218].


  
  
  

\bibitem{Krauss:1988zc} 
  L.~M.~Krauss and F.~Wilczek,
  ``Discrete Gauge Symmetry in Continuum Theories,''
  Phys.\ Rev.\ Lett.\  {\bf 62}, 1221 (1989).

\bibitem{Aulakh:1999cd}
  C.~S.~Aulakh, A.~Melfo, A.~Rasin and G.~Senjanovic,
  ``Seesaw and supersymmetry or exact R-parity,''
  Phys.\ Lett.\ B {\bf 459} (1999) 557
  [hep-ph/9902409].
 
\bibitem{Aulakh:2000sn}
  C.~S.~Aulakh, B.~Bajc, A.~Melfo, A.~Rasin and G.~Senjanovic,
  ``SO(10) theory of R-parity and neutrino mass,''
  Nucl.\ Phys.\ B {\bf 597} (2001) 89
  [hep-ph/0004031].

\bibitem{Babu:2008ep} 
  K.~S.~Babu and R.~N.~Mohapatra,
  ``Minimal Supersymmetric Left-Right Model,''
  Phys.\ Lett.\ B {\bf 668}, 404 (2008)
  [arXiv:0807.0481 [hep-ph]].

\bibitem{Feldman:2011ms}
  D.~Feldman, P.~Fileviez Perez and P.~Nath,
  ``R-parity Conservation via the Stueckelberg Mechanism: LHC and Dark Matter Signals,
''
  arXiv:1109.2901 [hep-ph].

\bibitem{FileviezPerez:2011dg} 
  P.~Fileviez Perez and M.~B.~Wise,
  ``Low Energy Supersymmetry with Baryon and Lepton Number Gauged,''
  Phys.\ Rev.\ D {\bf 84}, 055015 (2011)
  [arXiv:1105.3190 [hep-ph]].

\bibitem{Aulakh:1982yn} 
  C.~S.~Aulakh and R.~N.~Mohapatra,
  ``Neutrino as the Supersymmetric Partner of the Majoron,''
  Phys.\ Lett.\ B {\bf 119}, 136 (1982).

\bibitem{Hayashi:1984rd} 
  M.~J.~Hayashi and A.~Murayama,
  ``Radiative Breaking of $SU(2)_R X U(1)_{(B-L)}$ Gauge Symmetry Induced by Broken $N
=1$ Supergravity in a Left-right Symmetric Model,''
  Phys.\ Lett.\ B {\bf 153}, 251 (1985).

\bibitem{Mohapatra:1986aw}
  R.~N.~Mohapatra,
  ``Mechanism For Understanding Small Neutrino Mass In Superstring Theories,''
  Phys.\ Rev.\ Lett.\  {\bf 56} (1986) 561.

\bibitem{Masiero:1990uj}
  A.~Masiero and J.~W.~F.~Valle,
  ``A Model For Spontaneous R Parity Breaking,''
  Phys.\ Lett.\ B {\bf 251} (1990) 273.

\bibitem{Braun:2005zv} 
  V.~Braun, Y.~H.~He, B.~A.~Ovrut and T.~Pantev,
  ``Vector bundle extensions, sheaf cohomology, and the heterotic standard model,''
  Adv.\ Theor.\ Math.\ Phys.\  {\bf 10}, no. 4, 525 (2006)
  doi:10.4310/ATMP.2006.v10.n4.a3
  [hep-th/0505041].


\bibitem{Braun:2005nv} 
  V.~Braun, Y.~H.~He, B.~A.~Ovrut and T.~Pantev,
  ``The Exact MSSM spectrum from string theory,''
  JHEP {\bf 0605}, 043 (2006)
  doi:10.1088/1126-6708/2006/05/043
  [hep-th/0512177]. 
  
\bibitem{Ambroso:2009jd} 
  M.~Ambroso and B.~Ovrut,
  ``The B-L/Electroweak Hierarchy in Heterotic String and M-Theory,''
  JHEP {\bf 0910}, 011 (2009)
  doi:10.1088/1126-6708/2009/10/011
  [arXiv:0904.4509 [hep-th]].
  
\bibitem{Ambroso:2010pe} 
  M.~Ambroso and B.~A.~Ovrut,
  ``The Mass Spectra, Hierarchy and Cosmology of B-L MSSM Heterotic Compactifications,''
  Int.\ J.\ Mod.\ Phys.\ A {\bf 26}, 1569 (2011)
  doi:10.1142/S0217751X11052943
  [arXiv:1005.5392 [hep-th]].
  

\bibitem{Aad:2014aba} 
  G.~Aad {\it et al.}  [ATLAS Collaboration],
  ``Measurement of the Higgs boson mass from the $H\rightarrow \gamma\gamma$ and $H \rightarrow ZZ^{*} \rightarrow 4\ell$ channels with the ATLAS detector using 25 fb$^{-1}$ of $pp$ collision data,''
  Phys.\ Rev.\ D {\bf 90}, 052004 (2014)
  [arXiv:1406.3827 [hep-ex]].

\bibitem{Chatrchyan:2013lba} 
  S.~Chatrchyan {\it et al.}  [CMS Collaboration],
  ``Observation of a new boson with mass near 125 GeV in pp collisions at $\sqrt{s}$ = 7 and 8 TeV,''
  JHEP {\bf 1306}, 081 (2013)
  [arXiv:1303.4571 [hep-ex]].
  
\bibitem{Ovrut:2012wg} 
  B.~A.~Ovrut, A.~Purves and S.~Spinner,
  ``Wilson Lines and a Canonical Basis of SU(4) Heterotic Standard Models,''
  JHEP {\bf 1211}, 026 (2012)
  doi:10.1007/JHEP11(2012)026
  [arXiv:1203.1325 [hep-th]].
  
\bibitem{Marshall:2014kea} 
  Z.~Marshall, B.~A.~Ovrut, A.~Purves and S.~Spinner,
 ``Spontaneous R-parity Breaking, Stop LSP Decays and the Neutrino Mass Hierarchy,''
  Phys.\ Lett.\ B {\bf 732}, 325 (2014)
  doi:10.1016/j.physletb.2014.03.052
  [arXiv:1401.7989 [hep-ph]].
  
\bibitem{Marshall:2014cwa} 
  Z.~Marshall, B.~A.~Ovrut, A.~Purves and S.~Spinner,
  ``LSP Squark Decays at the LHC and the Neutrino Mass Hierarchy,''
  Phys.\ Rev.\ D {\bf 90}, no. 1, 015034 (2014)
  doi:10.1103/PhysRevD.90.015034
  [arXiv:1402.5434 [hep-ph]].
  
\bibitem{Ovrut:2014rba} 
  B.~A.~Ovrut, A.~Purves and S.~Spinner,
  ``A statistical analysis of the minimal SUSY BL theory,''
  Mod.\ Phys.\ Lett.\ A {\bf 30}, no. 18, 1550085 (2015)
  doi:10.1142/S0217732315500856
  [arXiv:1412.6103 [hep-ph]].
  
\bibitem{Ovrut:2015uea} 
  B.~A.~Ovrut, A.~Purves and S.~Spinner,
  ``The minimal SUSY $B-L$ model: from the unification scale to the LHC,''
  JHEP {\bf 1506}, 182 (2015)
  doi:10.1007/JHEP06(2015)182
  [arXiv:1503.01473 [hep-ph]].
  
  
  
  
\bibitem{Braun:2004xv} 
  V.~Braun, B.~A.~Ovrut, T.~Pantev and R.~Reinbacher,
  ``Elliptic Calabi-Yau threefolds with Z(3) x Z(3) Wilson lines,''
  JHEP {\bf 0412}, 062 (2004)
  doi:10.1088/1126-6708/2004/12/062
  [hep-th/0410055]. 

  
\bibitem{Donagi:2000zf} 
  R.~Donagi, B.~A.~Ovrut, T.~Pantev and D.~Waldram,
  ``Standard model bundles on nonsimply connected Calabi-Yau threefolds,''
  JHEP {\bf 0108}, 053 (2001)
  doi:10.1088/1126-6708/2001/08/053
  [hep-th/0008008].

\bibitem{FileviezPerez:2008sx}
  P.~Fileviez Perez and S.~Spinner,
  ``Spontaneous R-Parity Breaking and Left-Right Symmetry,''
  Phys.\ Lett.\ B {\bf 673} (2009) 251
  [arXiv:0811.3424 [hep-ph]].





\bibitem{Kaplunovsky:1992vs} 
  V.~S.~Kaplunovsky,
  ``One loop threshold effects in string unification,''
  hep-th/9205070.
  
  
\bibitem{Kaplunovsky:1995jw} 
  V.~Kaplunovsky and J.~Louis,
  ``On Gauge couplings in string theory,''
  Nucl.\ Phys.\ B {\bf 444}, 191 (1995)
  doi:10.1016/0550-3213(95)00172-O
  [hep-th/9502077].
  
  
\bibitem{Mayr:1993kn} 
  P.~Mayr, H.~P.~Nilles and S.~Stieberger,
  ``String unification and threshold corrections,''
  Phys.\ Lett.\ B {\bf 317}, 53 (1993)
  doi:10.1016/0370-2693(93)91569-9
  [hep-th/9307171].
  
  
\bibitem{Dienes:1995sq} 
  K.~R.~Dienes, A.~E.~Faraggi and J.~March-Russell,
  ``String unification, higher level gauge symmetries, and exotic hypercharge normalizations,''
  Nucl.\ Phys.\ B {\bf 467}, 44 (1996)
  doi:10.1016/0550-3213(96)00085-5
  [hep-th/9510223].
  
  
\bibitem{Dienes:1996du} 
  K.~R.~Dienes,
  ``String theory and the path to unification: A Review of recent developments,''
  Phys.\ Rept.\  {\bf 287}, 447 (1997)
  doi:10.1016/S0370-1573(97)00009-4
  [hep-th/9602045].
  
\bibitem{Dolan:1992nf} 
  L.~Dolan and J.~T.~Liu,
  ``Running gauge couplings and thresholds in the type II superstring,''
  Nucl.\ Phys.\ B {\bf 387}, 86 (1992)
  doi:10.1016/0550-3213(92)90047-F
  [hep-th/9205094].

  
\bibitem{Kiritsis:1996dn} 
  E.~Kiritsis, C.~Kounnas, P.~M.~Petropoulos and J.~Rizos,
  ``Universality properties of N=2 and N=1 heterotic threshold corrections,''
  Nucl.\ Phys.\ B {\bf 483}, 141 (1997)
  doi:10.1016/S0550-3213(96)00550-0
  [hep-th/9608034].
  
  
\bibitem{Nilles:1997vk} 
  H.~P.~Nilles and S.~Stieberger,
  ``String unification, universal one loop corrections and strongly coupled heterotic string theory,''
  Nucl.\ Phys.\ B {\bf 499}, 3 (1997)
  doi:10.1016/S0550-3213(97)00315-5
  [hep-th/9702110].
  
\bibitem{Nilles:1998uy} 
  H.~P.~Nilles,
  ``On the Low-energy limit of string and M theory,''
  [hep-ph/0004064].  
 

\bibitem{Ghilencea:2001qq} 
  D.~M.~Ghilencea and G.~G.~Ross,
  ``Precision prediction of gauge couplings and the profile of a string theory,''
  Nucl.\ Phys.\ B {\bf 606}, 101 (2001)
  doi:10.1016/S0550-3213(01)00234-6
  [hep-ph/0102306].
  
\bibitem{Klaput:2010dg} 
  M.~A.~Klaput and C.~Paleani,
  ``The computation of one-loop heterotic string threshold corrections for general orbifold models with discrete Wilson lines,''
  arXiv:1001.1480 [hep-th].
  

\bibitem{deAlwis:2012bm} 
  S.~P.~de Alwis,
  ``Gauge Threshold Corrections and Field Redefinitions,''
  Phys.\ Lett.\ B {\bf 722}, 176 (2013)
  doi:10.1016/j.physletb.2013.04.007
  [arXiv:1211.5460 [hep-th]].
  
\bibitem{Bailin:2014nna} 
  D.~Bailin and A.~Love,
  ``Reduced modular symmetries of threshold corrections and gauge coupling unification,''
  JHEP {\bf 1504}, 002 (2015)
  doi:10.1007/JHEP04(2015)002
  [arXiv:1412.7327 [hep-th]].
 


    

  
  
\bibitem{Lukas:1998qs} 
  A.~Lukas, B.~A.~Ovrut and D.~Waldram,
  ``Cosmological solutions of Horava-Witten theory,''  
  Phys.\ Rev.\ D {\bf 60}, 086001 (1999)
  doi:10.1103/PhysRevD.60.086001
  [hep-th/9806022].

\bibitem{inPreparation}
  R.~Deen, B.~A.~Ovrut, A.~Purves, in preparation.

\bibitem{Barger:2008wn} 
  V.~Barger, P.~Fileviez Perez and S.~Spinner,
  ``Minimal gauged U(1)(B-L) model with spontaneous R-parity violation,''
  Phys.\ Rev.\ Lett.\  {\bf 102}, 181802 (2009)
  doi:10.1103/PhysRevLett.102.181802
  [arXiv:0812.3661 [hep-ph]].
  
\bibitem{Everett:2009vy} 
  L.~L.~Everett, P.~Fileviez Perez and S.~Spinner,
  ``The Right Side of Tev Scale Spontaneous R-Parity Violation,''
  Phys.\ Rev.\ D {\bf 80}, 055007 (2009)
  doi:10.1103/PhysRevD.80.055007
  [arXiv:0906.4095 [hep-ph]].


\bibitem{Ghosh:2010hy}
  D.~K.~Ghosh, G.~Senjanovic and Y.~Zhang,
  ``Naturally Light Sterile Neutrinos from Theory of R-parity,''
  Phys.\ Lett.\ B {\bf 698} (2011) 420
  [arXiv:1010.3968 [hep-ph]].

\bibitem{Barger:2010iv}
  V.~Barger, P.~Fileviez Perez and S.~Spinner,
  ``Three Layers of Neutrinos,''
  Phys.\ Lett.\ B {\bf 696} (2011) 509
  [arXiv:1010.4023 [hep-ph]].

\bibitem{Mukhopadhyaya:1998xj} 
  B.~Mukhopadhyaya, S.~Roy and F.~Vissani,
  ``Correlation between neutrino oscillations and collider signals of supersymmetry in
 an R-parity violating model,''
  Phys.\ Lett.\ B {\bf 443}, 191 (1998)
  [hep-ph/9808265].

\bibitem{Chun:1998ub} 
  E.~J.~Chun and J.~S.~Lee,
  ``Implication of Super-Kamiokande data on R-parity violation,''
  Phys.\ Rev.\ D {\bf 60}, 075006 (1999)
  [hep-ph/9811201].

\bibitem{Chun:1999bq} 
  E.~J.~Chun and S.~K.~Kang,
  ``One loop corrected neutrino masses and mixing in supersymmetric standard model without R-parity,''
  Phys.\ Rev.\ D {\bf 61}, 075012 (2000)
  [hep-ph/9909429].

\bibitem{Hirsch:2000ef} 
  M.~Hirsch, M.~A.~Diaz, W.~Porod, J.~C.~Romao and J.~W.~F.~Valle,
  ``Neutrino masses and mixings from supersymmetry with bilinear R parity violation: A
 Theory for solar and atmospheric neutrino oscillations,''
  Phys.\ Rev.\ D {\bf 62}, 113008 (2000)
  [Erratum-ibid.\ D {\bf 65}, 119901 (2002)]
  [hep-ph/0004115].

\bibitem{FileviezPerez:2012mj}
  P.~Fileviez Perez and S.~Spinner,
  ``The Minimal Theory for R-parity Violation at the LHC,''
  arXiv:1201.5923 [hep-ph].

\bibitem{Perez:2013kla} 
  P.~Fileviez Perez and S.~Spinner,
  ``Supersymmetry at the LHC and The Theory of R-parity,''
  Phys.\ Lett.\ B {\bf 728}, 489 (2014)
  [arXiv:1308.0524 [hep-ph]].

\bibitem{Gamberini:1989jw}
  G.~Gamberini, G.~Ridolfi and F.~Zwirner,
  ``On Radiative Gauge Symmetry Breaking in the Minimal Supersymmetric Model,''
  Nucl.\ Phys.\  B {\bf 331}, 331 (1990).

\bibitem{PDG}
  K.A. Olive et al. (Particle Data Group),
  Chin. Phys. C, 38, 090001 (2014).

\bibitem{Djouadi:2005gi} 
  A.~Djouadi,
  ``The Anatomy of electro-weak symmetry breaking. I: The Higgs boson in the standard model,''
  Phys.\ Rept.\  {\bf 457}, 1 (2008)
  [hep-ph/0503172].

\bibitem{Porod:2000hv} 
  W.~Porod, M.~Hirsch, J.~Romao and J.~W.~F.~Valle,
  ``Testing neutrino mixing at future collider experiments,''
  Phys.\ Rev.\ D {\bf 63}, 115004 (2001)
  [hep-ph/0011248].

\bibitem{Hirsch:2003fe} 
  M.~Hirsch and W.~Porod,
  ``Neutrino properties and the decay of the lightest supersymmetric particle,''
  Phys.\ Rev.\ D {\bf 68}, 115007 (2003)
  [hep-ph/0307364].

\bibitem{Graham:2012th} 
  P.~W.~Graham, D.~E.~Kaplan, S.~Rajendran and P.~Saraswat,
  ``Displaced Supersymmetry,''
  JHEP {\bf 1207}, 149 (2012)
  [arXiv:1204.6038 [hep-ph]].


\bibitem{Graham:2014vya} 
  P.~W.~Graham, S.~Rajendran and P.~Saraswat,
  ``Supersymmetric crevices: Missing signatures of R -parity violation at the LHC,''
  Phys.\ Rev.\ D {\bf 90}, no. 7, 075005 (2014)
  [arXiv:1403.7197 [hep-ph]].



  
  
\bibitem{ArkaniHamed:2004fb} 
  N.~Arkani-Hamed and S.~Dimopoulos,
  ``Supersymmetric unification without low energy supersymmetry and signatures for fine-tuning at the LHC,''
  JHEP {\bf 0506}, 073 (2005)
  doi:10.1088/1126-6708/2005/06/073
  [hep-th/0405159].
  

  
\bibitem{Giudice:2004tc} 
  G.~F.~Giudice and A.~Romanino,
  ``Split supersymmetry,''
  Nucl.\ Phys.\ B {\bf 699}, 65 (2004)
  Erratum: [Nucl.\ Phys.\ B {\bf 706}, 487 (2005)]
  doi:10.1016/j.nuclphysb.2004.11.048, 10.1016/j.nuclphysb.2004.08.001
  [hep-ph/0406088].
  
  
\bibitem{ArkaniHamed:2004yi} 
  N.~Arkani-Hamed, S.~Dimopoulos, G.~F.~Giudice and A.~Romanino,
  ``Aspects of split supersymmetry,''
  Nucl.\ Phys.\ B {\bf 709}, 3 (2005)
  doi:10.1016/j.nuclphysb.2004.12.026
  [hep-ph/0409232].
  

\end{thebibliography}
\end{document}